\documentclass{aastex631}
\usepackage{float}
\usepackage{epsfig}
\usepackage{epstopdf}
\usepackage{amsmath}
\usepackage{hyperref}
\hypersetup{colorlinks=True,citecolor=blue}
\usepackage{amsfonts}
\usepackage{amssymb}
\usepackage{float}
\usepackage{bm}
\usepackage{graphicx}
\usepackage{color,xcolor}
\usepackage{verbatim}
\usepackage{subfigure}
\usepackage{graphicx}
\usepackage{multirow}
\usepackage{makecell}
\usepackage{listings}

\setcitestyle{authoryear,round}%Setting citation style. Changing braces to ()

\def\beq{\begin{equation}}
\def\eeq{\end{equation}}
\def\ber{\begin{eqnarray}}
\def\eer{\end{eqnarray}}
\def\benu{\begin{enumerate}}
\def\eenu{\end{enumerate}}

\def\l{\left}
\def\r{\right}
\def\d{{\rm d}}
\def\n{\nabla}

\def\f{\frac}

\def\l{\Lambda}
\def\b{{\rm b}}
\def\m{{\rm m}}

\newcommand{\omt}{\Omega_{0 \rm m}}
\newcommand{\omr}{\Omega_{0 \rm r}}
\newcommand{\omg}{\Omega_{0 \gamma}}

\newcommand{\oml}{\Omega_{\ell}}

\newcommand{\omlb}{\Omega_{\Lambda_{\rm b}}}

\newcommand{\omb}{\Omega_{0 \rm b}}
\def \lya {Ly$\alpha$}
\def \C {C_{\rm dr}}

\def \lleq {\lower0.9ex\hbox{ $\buildrel < \over \sim$} ~}
\def \ggeq {\lower0.9ex\hbox{ $\buildrel > \over \sim$} ~}

\def\n {\noindent}

\def \lcdm {$\Lambda$CDM\@}

\def \Olb {\Omega_{\Lambda_{\rm b}}}
\def \Oc {\Omega_{\rm dr}}
\def \weff {w_{\rm eff}}
\def \zd {z_d}
\def \rs {r_s^*}
\def \das {D_A(z_*)}
\def \ts {\theta_s}
\newcommand{\omdr}{\Oc}

\begin{document}

\title{Phantom braneworld and  the Hubble tension}
\author[0000-0003-0141-606X]{Satadru Bag}
\email{satadru@kasi.re.kr}
\affiliation{Korea Astronomy and Space Science Institute, Daejeon 34055, Korea}

\author[0000-0002-9470-9939]{Varun Sahni}
\email{varun@iucaa.in}
\affiliation{Inter-University Centre for Astronomy and Astrophysics,
Post Bag 4, Ganeshkhind, Pune 411~007, India}

\author[0000-0001-6815-0337]{Arman Shafieloo}
\email{shafieloo@kasi.re.kr}
\affiliation{Korea Astronomy and Space Science Institute, Daejeon 34055, Korea}
\affiliation{University of Science and Technology, Daejeon 34113, Korea}

\author[0000-0002-4891-7059]{Yuri Shtanov}
\email{shtanov@bitp.kiev.ua}
\affiliation{Bogolyubov Institute for Theoretical Physics,  Metrologichna St.~14-b, Kiev 03143, Ukraine}
\affiliation{Taras Shevchenko National University of Kiev, Volodymyrska St.~60, Kiev 01033, Ukraine}
% %\date{\today}

\begin{abstract}
Braneworld models with induced gravity exhibit phantom-like behaviour of the effective equation of state of dark energy. They can, therefore, naturally accommodate higher values of $H_0$, preferred by recent local measurements, while satisfying the CMB constraints. We test the background evolution in such phantom braneworld scenarios with the current observational datasets. We find that the phantom braneworld prefers a higher value of $H_0$ even without the R19 prior, thereby providing a much better fit to the local measurements. Although this braneworld model cannot fully satisfy all combinations of cosmological observables, among existing dark energy candidates the phantom brane provides one of the most compelling explanations of cosmic evolution.
\end{abstract}

%\keywords{Dark energy}
%\arxivnumber{....}

%\maketitle

\section{Introduction}
\label{sec:intro}

\n
Local measurements of the Hubble parameter $H_0$, based on a distance ladder treatment, appear to indicate a somewhat higher value of this quantity than that inferred from fitting $\Lambda$CDM to the cosmic microwave background
(CMB) data \citep{riess11A,linder13,riess16,bonvin,riess18,birrer,planck2016,planck2018}
(see, however, \citet{breuval19, Dainotti:2021pqg}).
This tension currently stands at about the $4.4\, \sigma$ level \citep{riess19}.

In the context of evolving dark energy, two different approaches have been advanced to minimize the $H_0$ tension (for a detailed review of all existing solutions, see \citet{DiValentino:2021izs}).
The first one focuses on the physics prior to recombination
\citep{early-kamion0,early-anjan18,early-kamion1,early-kamion2,early-randall,early-starobinsky,early-lin,early-lin1,early-smith,early-rishi,Ye:2020btb,Gogoi:2020qif,Seto:2021xua,Vagnozzi:2021gjh} while the second one
explores late-time physics, particularly
the impact of non-$\Lambda$CDM cosmologies
such as models with interacting dark sector and phantom behaviour 
\citep{late-silk,late-ujjaini,late-melchiorri1,late-arman1,late-kaplinghat1,late-arman2,Li:2019yem,Li:2020ybr,late-kaplinghat2,late-raveri19,late-shinji,late-vagnozzi,late-valentino,late-quintom,late-jailson} on the $H_0$ tension.
Our focus in the present paper will be on the second approach, i.e., models that exhibit phantom behaviour of dark energy.

CMB measurements determine the spacing between the acoustic peaks of the angular power spectrum, 
\beq
\theta_s = \frac{r_s^*}{D_A(z_*)}\;,
\label{eq:theta}
\eeq
very accurately at redshift $z_* \simeq 1100$.
Here, $r_{s}^*$ is the comoving sound horizon at the epoch of last scattering (we adopt the normalization $a_0 = 1$),
\beq
r_{s}^* = \int_0^{t_*}c_s(t)\frac{dt}{a(t)} 
= \int_{z_*}^\infty c_s(z)\frac{dz}{H(z)}\;,
\label{eq:rs}
\eeq
where $c_s$ is the sound speed, given by
\beq\label{eq:cs}
c_s(z)=\frac{c}{\sqrt{3} \sqrt{1 + \frac{3\omb}{4\omg (1+z)}}}\;,
\eeq
 and $D_A(z_*)$ is the comoving angular diameter distance to the
last scattering surface,
\beq
D_A(z_*) = \int_0^{z_*}\frac{dz}{H(z)}~.
\label{eq:DA}
\eeq

In a spatially flat $\Lambda$CDM cosmology, one has
\begin{equation} \label{H(z)}
H (z) = H_0 \left[ \Omega_{0 \rm m} ( 1 + z)^3 + \omr (1 + z)^4 + \Omega_\Lambda \right]^{1/2} \, ,
\end{equation}
where $\Omega_{0 \rm m}$ includes the contribution from dark and baryonic matter, and $\omr$ includes the contribution from photons as well as other possible relativistic matter such as neutrino's.

Although both $\rs$ and $\das$ explicitly contain $H_0$ in the denominator, a careful  examination reveals that they depend on $H_0$ differently (for a fixed $\omt$). This is because the parameter $\omr$ is determined in cosmology in the combination $\omr h^2$ (here, $h \equiv H_0/100\, {\rm km~s}^{-1}\, {\rm Mpc}^{-1}$), which makes $\omr$ substantially $H_0$-dependent, unlike the other two parameters $\omt$ and $\Omega_\Lambda$. As $H(z)$ increases with $z$, the primary contribution to \eqref{eq:rs} and \eqref{eq:DA} comes from redshifts near the lower limits of the respective integrals. Since the integration in \eqref{eq:DA} starts from $z=0$, one can neglect the contribution from the radiation term in $H(z)$ at low redshifts, thus $\das$ roughly scales as $ H_0^{-1}$. On the other hand, $\rs$ depends on $H_0$ differently as the integration in \eqref{eq:rs} starts from $z_* \simeq 1100$ (the value of $z_*$ is almost unaffected by any reasonable variation in $H_0$). At these high redshifts, one cannot neglect the contribution from the radiation part in (\ref{H(z)}),  which is substantially $H_0$-dependent, as explained above. Therefore, the Hubble parameter no longer scales as $H_0$; rather $H(z) \propto H_0^n$, where $0<n<1$ at redshifts $z \simeq 1100$.  Furthermore, $c_s$ depends on the ratio $\omb h^2/\omg h^2$. Since $\omg h^2$ is fixed from CMB measurements and one can assume independent constraint on $\omb h^2$ from, say, big bang nucleosynthesis (BBN) \cite{Cooke:2017cwo},\footnote{Note that CMB imposes even more stringent constraint on $\omb h^2$ as described below in section \ref{sec:cmb}.} $c_s$ is almost insensitive to $H_0$ (a mild dependency comes from the minute variation of $z_*$ with $H_0$). One can estimate numerically that $\rs$ roughly scales as $H_0^{-0.5}$, which results in $\theta_s \propto H_0^{0.5}$ approximately ( for a fixed $\omt$). Therefore, an increase in $H_0$ also raises the value of $\theta_s$.

Since $\theta_s$ is fixed from CMB measurements, one way of compensating the increment (due to a larger locally measured value of $H_0$) is to reduce $r_{s}^*$.
As demonstrated by (\ref{eq:rs}), the quantity $r_{s}^*$ can be reduced by slightly increasing $H(z)$ (by about $10 \%$) just
before recombination. Early dark energy does precisely this \citep{early-randall,knox19}; however, in these models, the mean constraint on the Hubble parameter does not shift much to the higher values when one considers combination of all datasets including the CMB polarisation data \citep{Hill:2020osr}.\footnote{A shift in $H_0$ can also be achieved by assuming the existence of primordial magnetic fields enhancing the recombination rate, thereby reducing the Hubble tension \citep{Jedamzik:2020krr}.}  

An alternative means of ameliorating the $H_0$ tension is to reduce the value of
$E(z) \equiv H(z)/H_0$ at lower redshifts, $z \ll z_*$.
This increases $\das$ in (\ref{eq:DA}) and therefore compensates for the increase in $\ts$ brought about by a larger value of $H_0$. This can be done more or less phenomenologically in models of evolving or dynamical dark energy \citep{DiValentino:2021izs}.
It is interesting in this respect that a smaller value of $E(z)$ at low/medium $z$ (compared to that of $\Lambda$CDM model) is a {\em generic\/} (and quite incidental) feature of a certain type of braneworld cosmology. Specifically, the model with one extra dimension and with induced gravity on the brane has two branches of cosmological solutions, one of which (so-called `normal branch') exhibits this property \citep{ss02, loiter, mimicry}.
On this branch, the effective equation of state (EoS) of dark energy is
phantom-like, $w_{\rm eff} < -1$, for which reason it was also termed `phantom brane' \citep{Bag:2016tvc}. One can envisage that the $H_0$ tension will be alleviated in this braneworld scenario. In this work, we focus on the phantom brane, and test whether it can indeed alleviate the $H_0$ tension.

The phantom brane model has been tested using distance measures in the literature; see, e.g., \citet{Lazkoz:2006gp, late-ujjaini}. However, in view of the many interesting features of this model, including the potential of reducing the $H_0$ tension, we re-examine 
the braneworld scenarios in the light of the newer data sets. In this work, we also relax the simplifying assumption of zero bulk cosmological constant, made in the previous studies.

Our paper is organized as follows. Section \ref{sec:BW} describes the braneworld models with phantom behaviour in detail. In section \ref{sec:testing}, we test these models against the observation of the background evolution and compare the results with that in the $\Lambda$CDM model, focusing on the parameter $H_0$ in particular. We present our conclusions in section \ref{sec:con}.
\section{Braneworld Cosmology}
\label{sec:BW}

Braneworld models are effective field-theoretic models with large (often noncompact) extra dimensions. Motivated by the studies of relations between string theory and supergravity \citep{Horava:1995qa}, they were subsequently used as theoretical schemes for addressing the Planck hierarchy problem in \citet{Antoniadis:1998ig, Arkani-Hamed:1998jmv, Randall:1999ee, Randall:1999vf}.  Their application to cosmology opened up a new area of cosmological investigations 
(see \citet{Maartens:2010ar} for a review).

One popular version that attracted much attention in cosmology is the braneworld model with a single large extra dimension and with `induced gravity' on the brane. It is described by the action \citep{Dvali:2000hr, Collins:2000yb, Shtanov:2000vr}
\begin{equation}\label{action}
S = M^3 \int_{\rm bulk} \left( {\cal R} - 2 \Lambda_{\rm b} \right) + m^2 \int_{\rm brane} \left( R - 2 \Lambda \right) + \int_{\rm brane} L_{\rm m} \, . 
\end{equation} 
This model represents a simple general-relativistic action in the five-dimensional bulk (with scalar curvature ${\cal R}$) and on the four-dimensional brane (with scalar curvature $R$), with matter confined only to the brane and described by the Lagrangian $L_{\rm m}$. Integrations over the bulk and
brane are taken with the corresponding natural volume elements. The universal constants $m$ and $M$ play the role of the Planck masses on the brane and in the bulk space, respectively. The symbols $\Lambda$ and  $\Lambda_{\rm b}$ denote, respectively, the cosmological constants on the brane and in the bulk, so that $m^2 \Lambda$ is the brane tension from the five-dimensional bulk perspective.  

Braneworld cosmology is further specified by the presence or absence of a black hole or naked singularity in the bulk space. Within this setting, the cosmological equation on the brane with total energy density $\rho$ has the form \citet{ss02}
\begin{equation}\label{cosmob}
H^2 + \frac{\kappa}{a^2} = \frac{\rho}{3 m^2} + \frac{\Lambda}{3} + \frac{2}{\ell^2} \left[ 1 \pm \sqrt{1 + \ell^2 \left( \frac{\rho}{3 m^2} + \frac{\Lambda}{3} - \frac{\Lambda_{\rm b}}{6} - \frac{\C}{a^4} \right) } \right] \, , 
\end{equation}
where $\kappa = 0, \pm 1$ describes the spatial curvature, and we have introduced the characteristic length scale 
\begin{equation} \label{ell}
\ell = {2 m^2 \over M^3} \, .
\end{equation}
The term $\C / a^4$ under the square root in (\ref{cosmob}) reflects the presence of a black hole or a naked singularity in the bulk space, with mass proportional to the constant $\C$.

The ``$\pm$'' sign in (\ref{cosmob}) corresponds to two distinct ways of embedding the brane in the bulk geometry \citep{Collins:2000yb, Deffayet:2000uy}. In \citet{ss02} models with the lower (``$-$'') sign were referred to as Brane\,1, while models with the upper (``$+$'') sign were called Brane\,2. Later it was found that the Brane\,2 (also known as the self-accelerating branch) is plagued by the existence of {\em ghost excitations} \citep{Charmousis:2006pn,Gorbunov:2005zk,Koyama:2007za}. Therefore, in this article, we focus on the Brane\,1 (also known as the normal branch) which is a physically viable model of dark energy in this braneworld framework. It is this branch that generically exhibits phantom-like behaviour of dark energy and, for this reason, we also call it `phantom brane' \citep{Bag:2016tvc}.

As the mass $M$ tends to zero [or, equivalently, the length scale (\ref{ell}) tends to infinity], one smoothly recovers general relativity. Therefore, this parameter should always be non-zero in braneworld cosmology. As regards the bulk cosmological constant $\Lambda_{\rm b}$ or dark-radiation term in (\ref{cosmob}), they may well be equal to zero. Such a braneworld model with $\Lambda_{\rm b} = 0$ and $\C = 0$ is the simplest version of the phantom brane; we will call it {\em minimal phantom brane\/} in this paper. In this case, the brane is embedded in the flat bulk space. Compared to the $\Lambda$CDM model, the mimimal brane has one additional parameter, the length scale (\ref{ell}). A general model with no restriction on the constants will then be called {\em general phantom brane\/}. Compared to the minimal brane, it has two more free parameters, $\Lambda_{\rm b}$ and $\C$.

\subsection{Minimal phantom brane}\label{sec:minimal}

One can present the expansion history (\ref{cosmob}) on a spatially flat ($\kappa = 0$) minimal phantom brane in terms of cosmological redshift as follows \citep{ss02}:
\begin{equation} \label{hubble0}
E^2(z) \equiv \left[ \frac{H(z)}{H_0} \right]^2 = \omt (1\!+\!z)^3 + \Omega_\Lambda + 2 \Omega_\ell - 2 \sqrt{\Omega_\ell}\, \sqrt{\omt (1\!+\!z )^3 + \Omega_\Lambda + \Omega_\ell } \, ,
\end{equation}
where
\begin{equation} \label{omegas}
\omt =  {\rho_0 \over 3 m^2 H_0^2} \, , \qquad \Omega_\Lambda = {\Lambda
\over 3 H_0^2} \, , \qquad \Omega_\ell = {1 \over \ell^2 H_0^2} \, .
\end{equation}
We consider here only the late-time evolution, so the contribution from radiation to (\ref{hubble0}) is neglected.

The $\Omega$ parameters satisfy the constraint equation
\begin{equation} \label{omega-r1}
\omt + \Omega_\Lambda - 2 \sqrt{\Omega_\ell}
= 1 \, ,
\end{equation}
which reduces the number of independent $\Omega$ parameters.  

A key new ingredient in braneworld models compared to $\Lambda$CDM is the parameter $\Omega_\ell$, which encodes the presence of
a large (non-compact) extra dimension. In the limit of $\Omega_\ell \to 0$, equations
(\ref{hubble0}) and (\ref{omega-r1}) reduce to the $\Lambda$CDM limit
\begin{align} \label{hubL}
&E^2(z) = \omt (1\!+\!z)^3 + \Omega_\Lambda \, , \\
&\omt + \Omega_\Lambda = 1 \, ,
\label{LCDM}
\end{align}
with $\Omega_\Lambda$ corresponding to the usual cosmological constant.

As pointed out in \citet{ss02, Lue:2004za}, all phantom brane models imitate cosmologies with a phantom EoS of dark energy, $w_{\rm eff} < -1$.  Here, the effective equation of state is defined as \citep{ss02, Sahni:2006pa}
\begin{equation}\label{EoS}
w_{\rm eff} (z) = \frac{2 q (z) - 1}{3\left[ 1 - \Omega_{\rm m} (z) \right]} = \frac{2 (1 + z) E'(z) / E(z) - 1}{3 \left[ 1 - \Omega_{0 \rm m} (1 + z)^3 / E^2 (z)\right]} \, ,
\end{equation}
where $q(z)$ is the deceleration parameter. In the minimal phantom brane model, the effective EoS at the present epoch ($z=0$) has the form
\beq\label{eq:weff_bw_0}
w_0 \equiv w_{\rm eff}(z=0)=-1-\dfrac{\Omega_{0 \rm m}}{1-\Omega_{0 \rm m}} \left(\dfrac{\sqrt{\Omega_{\ell}}}{1+\sqrt{\Omega_{\ell}}}\right) \, ,
\eeq
demonstrating the phantom-like property $w_0 < -1$ of the effective dark energy.  

\begin{figure}
\includegraphics[width=0.328\textwidth]{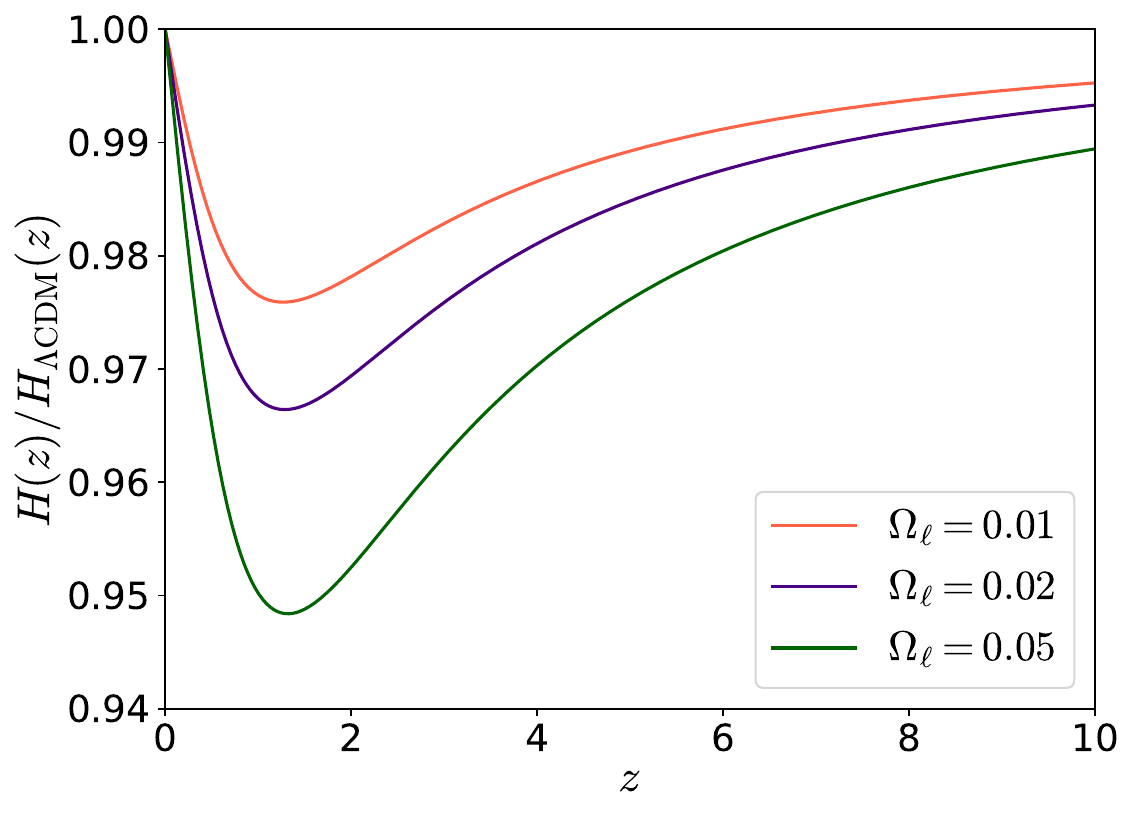}
\includegraphics[width=0.328\textwidth]{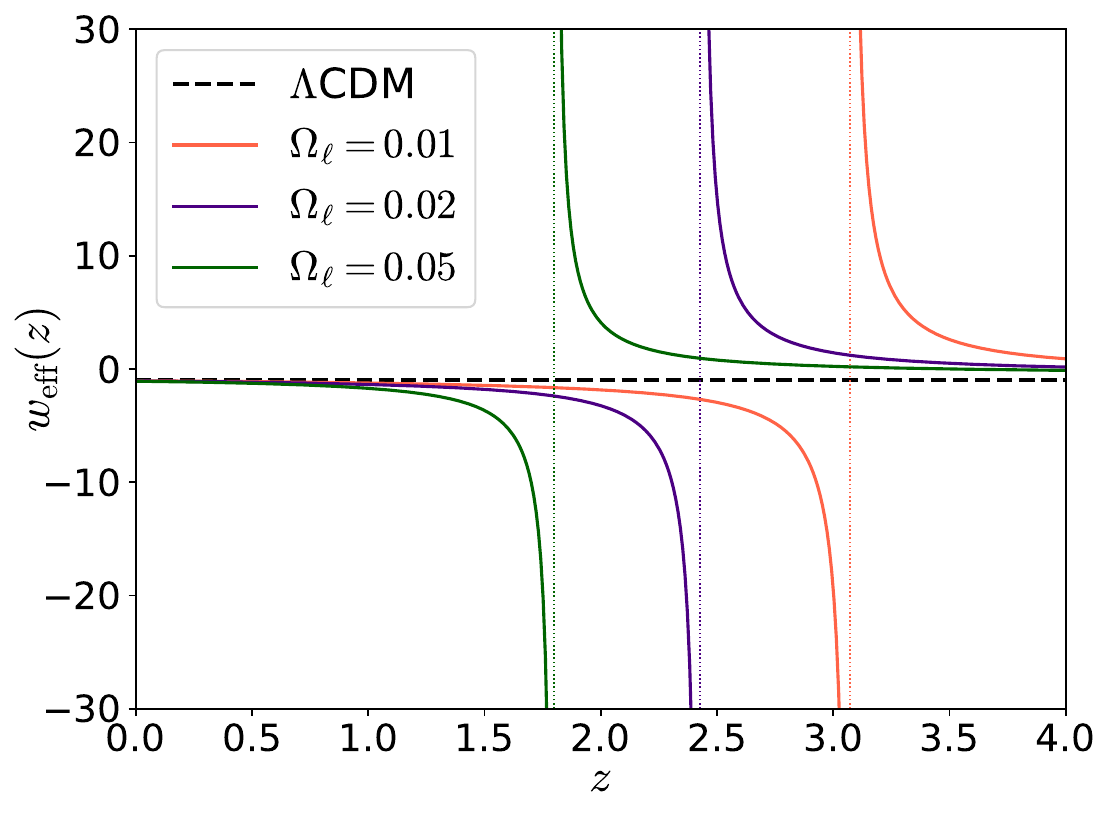}
\includegraphics[width=0.328\textwidth]{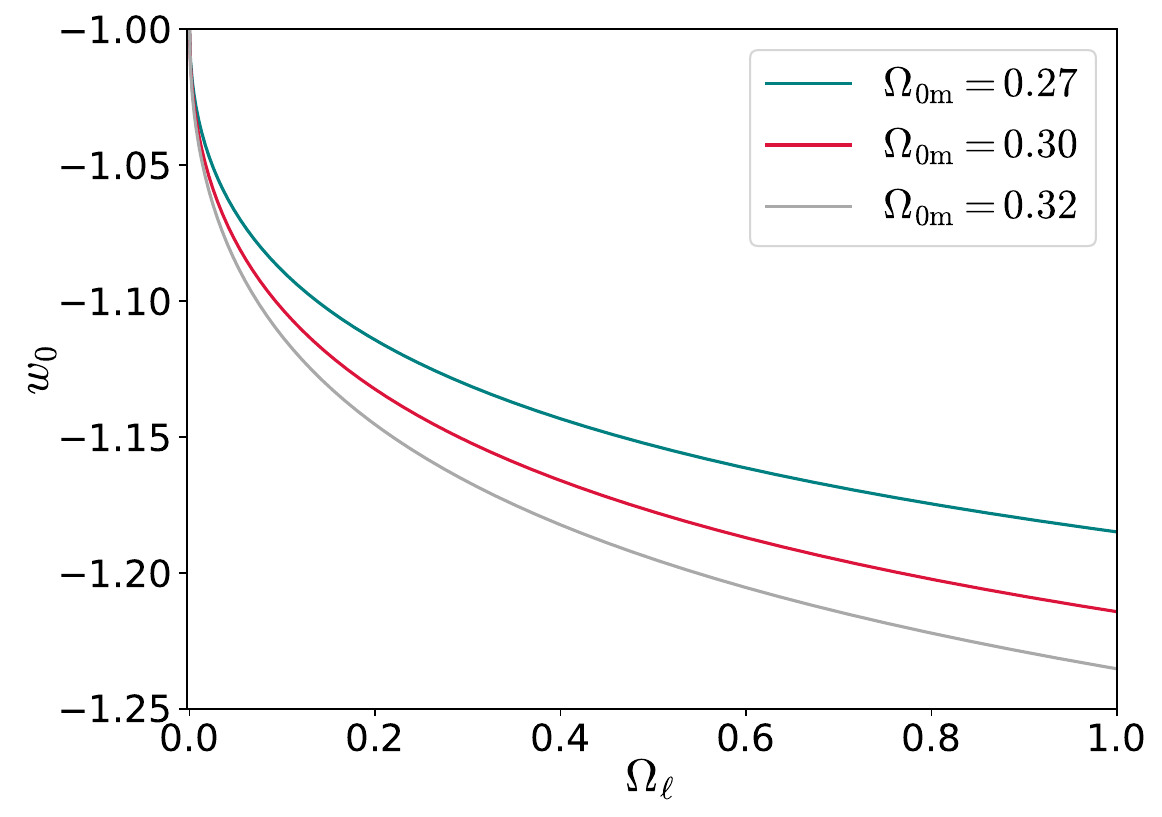} 
\caption{The left panel shows that the expansion rate is slower on the minimal phantom brane as compared to the $\Lambda$CDM model ($\omt=0.3$ is assumed); the larger is the value of the brane parameter $\oml$, the slower is the expansion rate. Evolution of the effective equation of state of dark energy on the minimal phantom brane is illustrated in the middle panel. For any $\oml>0$, $\weff(z)$ exhibits a pole at $z_p$ (shown by the dotted vertical lines) and the EoS behaves like phantom for $z<z_p$.  The right panel shows the present EoS ($w_0$)  as a function of $\Omega_{\ell}$ for different values of $\omt$. As $\oml \rightarrow 0$, one recovers the $\Lambda$CDM limit.}
\label{fig:background}
\end{figure}

The background evolution on the minimal phantom brane is shown in figure \ref{fig:background}. The left panel demonstrates that $H(z)$ on the brane is lower than that in $\Lambda$CDM\@. For larger values of the brane parameter $\oml$, the expansion rate on the phantom brane becomes slower. The middle panel displays the evolution of the effective equation of state (EoS) of the dark energy $w_{\rm eff}(z)$.  For any $\oml>0$, $\weff(z)$ exhibits a pole at $z_p$ (shown by the dotted vertical lines) and the EoS behaves like a phantom for $z<z_p$. The pole does not represent anything unphysical; it is the artefact of describing the braneworld evolution (\ref{hubble0}) in terms of an effective dark energy within the general relativity framework. The right panel shows the phantom-like present-epoch EoS ($w_0$) as a function of $\oml$ for three values of $\omt$. One recovers the $\Lambda$CDM limit as $\oml \rightarrow 0$.

The slower-than-$\Lambda$CDM value of the expansion rate at late times in phantom brane models affects several key observational quantities, including: 

\begin{itemize}

\item
The comoving angular-diameter distance to the
last scattering surface, $D_A(z_*)$, defined in \eqref{eq:DA}.
Since $H_{\rm brane} < H_{\Lambda \rm CDM}$, it follows that $D_A(z_*)$ is larger on
the phantom brane than in $\Lambda$CDM\@.
As briefly discussed in the introduction, this could alleviate the $H_0$ tension existing in
$\Lambda$CDM\@. Indeed, tests of the minimal phantom brane model with earlier datasets
indicate that larger values of $H_0$ are compatible with a reasonable choice of the brane parameters, as discussed in \citet{late-ujjaini}.

\item
The age of the universe
\beq\label{eq:age}
t(z) = \int_z^\infty \frac{dz'}{(1+z') H(z')} \, .
\eeq
Since $H_{\rm brane} < H_{\Lambda \rm CDM}$, the phantom brane has a larger age
than $\Lambda$CDM.

A larger age of the universe makes the presence of high-redshift objects such as galaxies and QSO's easier to account for in the phantom brane relative to $\Lambda$CDM\@. The recent study \citet{Vagnozzi:2021tjv} shows that, indeed, it is difficult to accommodate the ages of high-redshift galaxies and QSO's in the \lcdm~ model.
Note that a version of the phantom brane model has been extensively used in $N$-body simulations to demonstrate the effect of the presence of a space-like large extra dimension on large-scale structure of the Universe \citep{Schmidt:2009sv,Bag:2018jle, Fiorini:2021dzs}.  

\end{itemize}

\subsection{General phantom brane}
\label{sec:general}

A general phantom brane has nonzero bulk cosmological constant $\Lambda_{\rm b}$ and possible dark-radiation constant $\C$ in (\ref{cosmob}). In this case, the evolution equation (\ref{hubble0}) generalizes to
\begin{equation} \label{hubble_loiter}
{H^2(z) \over H_0^2} = \omt (1\!+\!z)^3 + \Omega_\Lambda + 2
\Omega_\ell - 2 \sqrt{\Omega_\ell}\, \sqrt{\omt
(1\!+\!z )^3 + \Omega_\Lambda + \Omega_\ell + \Omega_{\Lambda_{\rm b}} +
\omdr(1\!+\!z )^4} \, ,
\end{equation}
and, in addition to (\ref{omegas}), contains two more parameters
\begin{equation} \label{omegaC}
\Omega_{\Lambda_{\rm b}} = - {\Lambda_{\rm b} \over 6 H_0^2} \, , \qquad \omdr= -\frac{\C}{a_0^4H_0^2} \, .
\end{equation}
The extended set of $\Omega$'s now satisfies the constraint equation
\begin{equation} \label{omega-r2}
\omt + \Omega_\Lambda - 2 \sqrt{\Omega_\ell}\, \sqrt{1 +
\Omega_{\Lambda_{\rm b}} + \omdr} = 1 \, .
\end{equation}
Equations (\ref{hubble_loiter}), (\ref{omega-r2}) reduce to (\ref{hubble0}), (\ref{omega-r1}) when dark radiation and bulk cosmological constant are absent ($\Omega_{\Lambda_{\rm b}} = \Oc =0$), and to the $\Lambda$CDM equations (\ref{hubL}), (\ref{LCDM}) when $\Omega_\ell = 0$. Remarkably, in the scenarios where the dark-radiation term dominates over the other terms inside the square-root in \eqref{hubble_loiter}, the braneworld mimics the expansion of a spatially closed $\Lambda$CDM model at late times \citep{loiter}.

\section{Testing against observations}
\label{sec:testing}

\subsection{Datasets}
To test the background evolution of the braneworld models, we consider the observations of type-1a supernovae (SNe), baryon acoustic oscillation (BAO) and comic microwave background expressed in terms of background parameters. The datasets are described below.

\subsubsection{JLA supernovae compilation}
We use the JLA compilation \citep{Betoule:2014frx} of $740$ type-Ia SNe distributed over $z \in [ 0.01006,1.299106 ]$.
The distance modulus is defined as
\beq \label{dmod}
\mu(z)=5 \times \log_{10} \left[d_L \times 10^5\right]\, ,
\eeq
where the luminosity distance,
\begin{equation}
    d_L(z)=\frac{c(1+z)}{H_0} \int_0^z \frac{dz'}{E(z')}\;,
\end{equation}
is given in Mpc. From the JLA data one can obtain the distance modulus
\beq
\mu=m^*_B - \left( M_B-\alpha X_1 + \beta C \right) \, ,
\eeq
where $m^*_B$ is the observed peak (apparent) magnitude in B-band of the SNe and $X_1$, $C$ are the stretch and the colour of the SNe, respectively.
The dependence of the absolute magnitude $M_B$ on the host galaxy mass is expressed by the condition
\begin{equation}\label{eqn:M_B}
  M_B = \left\{\begin{array}{rl}
        M_B^1, &\quad \text{if } M_{\mathrm{stellar}}<10^{10}M_{\mathrm{sun}} \, , \\
        M_B^1 +\Delta M, &\quad \text{otherwise } \, . 
        \end{array} \right. 
\end{equation}

In summary, we have four hyper-parameters, namely $M_B^1$, $\Delta M$, $X_1$, and $C$, which should be marginalised over while doing cosmology. Note that there is a degeneracy between $M_B^1$ and $H_0$; this degeneracy can be broken only when JLA dataset (or any SN dataset) is used together with other probes like BAO and/or CMB measurements.

\subsubsection{BAO: eBOSS DR16 compilation + MGS+ 6dfGS }
\begin{figure}
\includegraphics[width=0.49\textwidth]{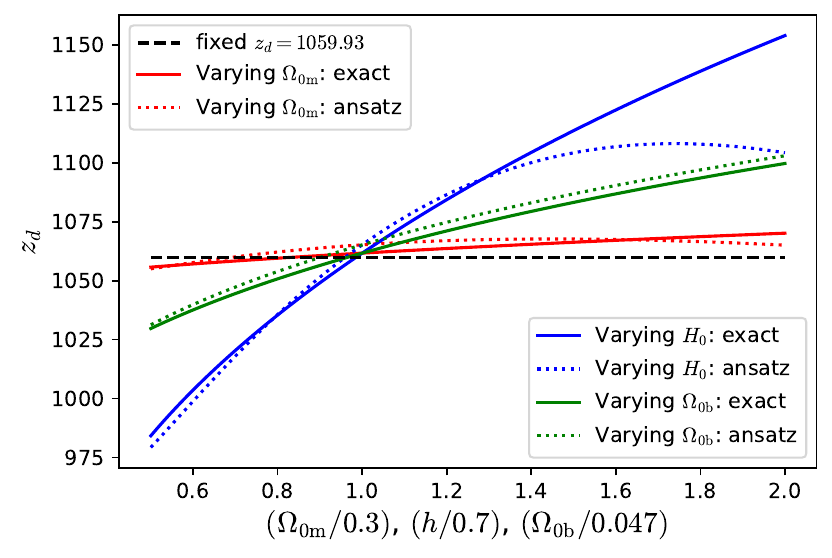}
\includegraphics[width=0.49\textwidth]{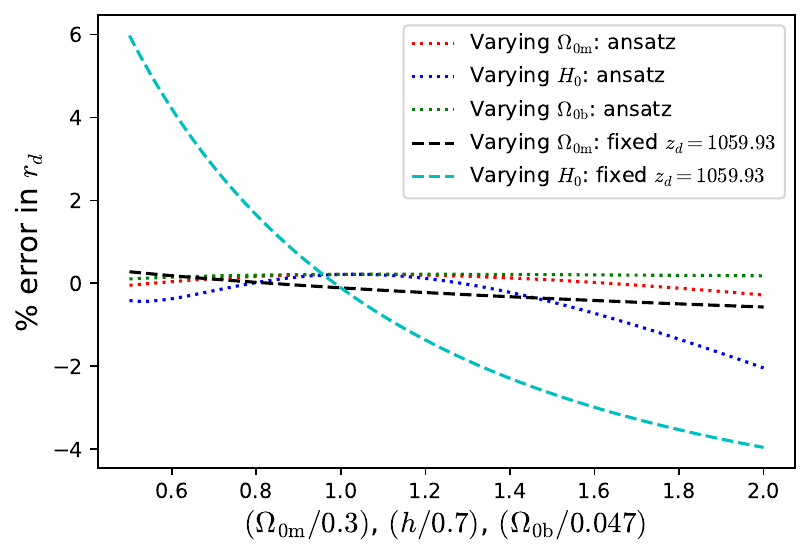} 
\caption{The left panel shows the change in $z_d$ with the variation of $\lbrace \omt, H_0, \omb \rbrace$ in the $\Lambda$CDM model by the blue, green and red curves, respectively. The solid curves represent the exact calculation by solving the Boltzmann equation using \lstinline{CAMB} whereas the dotted lines show the estimations using the ansatz \eqref{eq:zd_an}. The black dashed line shows the fixed $z_d=1059.93 \pm 0.46$ 
obtained from best fit {\em Planck\/} 2018 TT+TE+EE+LowE \citep{planck2018}. The right panel shows the percentage error in $r_d$ calculated using \eqref{eq:rd} with the ansatz for $z_d$ (dotted curves) and with the fixed $z_d$ (dashed curves).}
\label{fig:zd}
\end{figure}
We use the baryon acoustic oscillations (BAO) data from the eBOSS DR16 compilation (BOSS DR12\,+\,eBOSS Dr16\,+\,\lya) \citep{sdss12, eBOSS:2020yzd} together with 6dF galaxy survey (6dFGS) \citep{6df} and main galaxy sample (MGS) \citep{mgs2}. These data (summarised in table \ref{tab:bao}) are typically expressed in terms of the ratios of distances, $D_M/r_d$, $D_H/r_d$ and $D_V/r_d$, which are defined as follows:
\begin{align}
D_H(z) &= \frac{c}{H_0 E(z)}\;, \\
D_M(z) &=\frac{c}{H_0}\int_0^z \frac{dz'}{E(z')}\;, \\
D_V(z) &=\left[z D_H(z) D_M(z)^2 \right]^{1/3}\;.
\end{align}

The comoving sound horizon at the baryon drag epoch is given by
\beq\label{eq:rd}
r_d=\frac{1}{H_0}\int_{z_d}^\infty \frac{c_s(z)}{E(z)}dz\;,
\eeq
where $z_d$ is the redshift of the baryon drag epoch, and the sound speed $c_s (z)$ is given by (\ref{eq:cs}).

Since the phantom brane models we consider in this work do not modify the background evolution as well as the growth of perturbation compared to that in the GR for $z> \text{a few}$, the high-redshift baryonic physics remains unaltered in these models as compared to $\Lambda$CDM model. Thus, we can safely assume the value of the redshift of the baryon drag epoch ($z_d$) to be the same as in $\Lambda$CDM model.
One could use a fixed  $z_d$ or estimate it from ansatzes like the following one from \citet{{Hu:1995en}}:
\beq\label{eq:zd_an}
\zd = 1345 ~\frac{\left(\omt h^2\right)^{0.251}}{1+0.659 \left(\omt h^2\right)^{0.828}}~\left[1+b_1 \left(\omb h^2\right)^{b_2} \right]\;,
\eeq
where 
\begin{align}
b_1 &= 0.313~\left(\omt h^2\right)^{-0.419}~\left[ 1+0.607 \left(\omt h^2\right)^{0.674} \right]\;, \\ 
b_2 &= 0.238 \left(\omt h^2\right)^{0.223} \, .
\end{align}
In figure \ref{fig:zd}, we inspect how the above ansatz performs in $\Lambda$CDM model. 
In the left panel, we show the change in $z_d$ in $\Lambda$CDM model, calculated  using the ansatz \eqref{eq:zd_an}, with the variation of $\lbrace \omt, H_0, \omb \rbrace$. The exact value is what we obtain from \lstinline{CAMB}\footnote{\href{https://camb.info}{https://camb.info}} \citep{Lewis:1999bs} (by solving the Boltzmann equation). The right panel shows the percentage error in $r_d$ arising from the ansatz \eqref{eq:zd_an} for $z_d$ (dotted curves) and a fixed $z_d$ (dashed curves). From both panels, we have found that the fixed $z_d$ performs poorly with variation in $H_0$. On the other hand, the ansatz \eqref{eq:zd_an} provides a good match to the exact $z_d$ causing the sub-percent error in $r_d$ when $\lbrace \omt, H_0, \omb \rbrace$ are in a reasonable range. However, still we use the value of $z_d$ obtained from \lstinline{CAMB} for each point in the parameter space while sampling. We then calculate $r_d$ using \eqref{eq:rd}. Note that, for similar reasons, we use \lstinline{CAMB} again to calculate the redshift to last-scattering surface ($z_*$) while testing with CMB data.

\begin{table}
\renewcommand{\arraystretch}{2}
\setlength{\tabcolsep}{10pt}
\begin{tabular}{||c||c|c|c|c||}
\hline
Source & $z_{\rm eff}$ & $D_V/r_d$ & $D_M/r_d$ & $D_H/r_d$ \\
\hline
\hline

6dFGS & $0.106$& $ 3.047 \pm  0.137 $ & --- & --- \\
\hline

 MGS  & $ 0.15 $ & $ 4.508 \pm 0.135 $ & --- & --- \\
\hline
\hline

 BOSS DR12 LRG   & $ 0.38 $ & --- & $10.234 \pm 0.169   $ & $24.981 \pm 0.729 $ \\
\hline

 BOSS DR12 LRG   & $ 0.51 $ & --- & $ 13.366 \pm 0.204  $ & $  22.317 \pm  0.572$ \\
\hline
\hline

 eBOSS LRG & $  0.698 $ & --- & $ 17.858 \pm 0.328 $ & $ 19.326 \pm 0.533 $ \\
\hline

 eBOSS ELG  & $ 0.845  $ & $ {18.33}^{+0.57}_{-0.62}$ & --- & --- \\

\hline

eBOSS Quasar  & $  1.48 $ & --- & $ 30.688 \pm 0.798 $ & $ 13.261 \pm  0.552 $ \\
\hline
\hline
\lya-\lya  & $  2.334 $ & --- & $ 37.6\pm 1.9 $ & $ 8.93\pm 0.28  $ \\
\hline
 \lya-QSO & $ 2.334  $ & --- & $ 37.3 \pm 1.7 $ & $  9.08\pm 0.34 $ \\
\hline
\hline
\end{tabular}
\caption{BAO data from different surveys, collected from \citet{eBOSS:2020yzd}. Note that we use the covariance matrices for BOSS DR12 LRG, eBOSS DR16 LRG and eBOSS Quasar data whereas we compute the log probability directly from the likelihood tables provided for eBOSS DR16 ELG, \lya-\lya~ and \lya-QSO data. For the rest, i.e., for 6dFGS and MGS, we consider the individual datapoints as given above.}
\label{tab:bao}
\end{table}

\subsubsection{CMB reduced parameters}\label{sec:cmb}
Although we understand how cosmological perturbations grow on the minimal phantom brane \citep{Koyama:2005kd, Maartens:2010ar, Sawicki:2006jj, Seahra:2010fj, Mukohyama:2000ui, Mukohyama:2001yp, Viznyuk:2013ywa, Bag:2016tvc,Viznyuk:2018eiz}, we are not completely sure about the case with large contributions from dark radiation ($\Oc$) and bulk cosmological constant ($\Olb$). Therefore, we do not consider the full CMB angular power-spectrum data in this work. Rather, we use the CMB data expressed in terms of distances, precisely the the three `reduced' parameters $R(z_*)$, $l_A(z_*)$ and $\omega_{\rm b}$, defined as follows:
\begin{align} 
R &=\frac{\das}{c}\sqrt{\omt H_0^2}\;, \label{eq:R} \\
l_A(z_*) &= \frac{\pi \das}{\rs} \;, \label{eq:lA} \\
\omega_{\rm b} &=\omb h^2\;.
\end{align}
Here, $\rs$ and $\das$, defined in \eqref{eq:rs} and \eqref{eq:DA}, respectively, are the comoving sound horizon at the last-scattering surface (LSS) and the comoving angular diameter distance to the LSS at the redshift $z_*$. 
Again, one can use ansatzes like the one given in \citet{Hu:1995en} to calculate $z_*$. However, instead of using an ansatz for $z_*$ or a fixed $z_*$, we calculate it using \lstinline{CAMB} in $\Lambda$CDM model during sampling for better accuracy 
as for $z_d$ in the BAO case.

\begin{table}[!htp]
\centering
\renewcommand{\arraystretch}{2}
\setlength{\tabcolsep}{10pt}
\begin{tabular}{ccccc}
\hline\hline
  Parameters & {\em Planck\/} TT, TE, EE\,+\,lowE & $R$ & $l_\textrm{A}$ & $\omega_{\rm b}$ \\
\hline 
$R$    & $1.7493^{+0.0046}_{-0.0047}$ & $1.0$&    $0.47$&   $-0.66$\\
$l_\textrm{A}$ & $301.462^{+0.089}_{-0.090}$ & $0.47$&    $1.0$&   $-0.34$\\
$\omega_{\rm b}$ & $0.02239\pm0.00015$ & $-0.66$& $-0.34$&  $1.0$\\
\hline\hline
\end{tabular}
\caption{The $68~\%$ C.L. limits on the CMB reduced parameters $R$, $l_\textrm{A}$, $\omega_{\rm b}$ in $w$CDM model and their correlation matrix for {\em Planck\/} 2018 TT, TE, EE\,+\,lowE, taken from \citet{Chen:2018dbv}.}
\label{tab:CMB_data_wcdm}
\end{table}

From \citet{Chen:2018dbv}, we use the the fit values and the covariance matrix of these parameters, $\lbrace R, l_A, \omega_{\rm b} \rbrace$, in the $w$CDM model for {\em Planck\/}  2018 TT, TE, EE\,+\,lowE as the data which have been summarised in table \ref{tab:CMB_data_wcdm}\@. From the correlations one can get the inverse covariance matrix as
\beq \label{cov}
C^{-1}=\begin{pmatrix}
91232.8 & -1323.1 & 1616785.2 \\
-1323.1 & 158.8 & 5030.1 \\
1616785.2 & 5030.1 & 78905711.2 
\end{pmatrix}\;.
\eeq

\subsubsection{Direct $H_0$ measurement by Riess et al 2019 (R19 hereafter)}
We also consider the direct $H_0$ measurement by \citet{riess19}. However, we show the fit results both with and without this datapoint separately.

\subsection{Models considered}
We test two versions of the phantom brane model together with the base $\Lambda$CDM model, for comparison, on the datasets described above. The models are summarised below. Note that, following standard cosmology,  we consider three neutrino species, only one having mass $m_{\nu}=0.06$ eV, in the background with $N_{\rm eff}=3.046$ in all the models (nonzero neutrino mass requires a slightly lower value of $H_0$ in all the models).

\subsubsection{$\Lambda$CDM}
For testing at the background level, we need three free parameters ($\omt$, $H_0$ and $\omb$)  in the $\Lambda$CDM model, along with the four hyper-parameters needed for fitting the JLA dataset.

\subsubsection{Minimal phantom brane}

For the minimal phantom brane model, described in section \ref{sec:minimal}, we have one extra brane parameter relative to the  $\Lambda$CDM model, leading to a total of 
four free cosmological parameters: $\omt$, $H_0$, $\omb$ and the brane parameter $\oml\geq0$. Note that we also have the same four hyperparameters as in $\Lambda$CDM model to fit the JLA datasets.

\subsubsection{General phantom brane}

Next we consider the general phantom brane model, with nonzero bulk cosmological constant and dark radiation, as explained in section \ref{sec:general}\@.
Thus, we now have six free cosmological parameters:  $\omt$, $H_0$, $\omb$, $\oml$, $\omlb$ and $\Oc$, on which we set priors $\oml>0$ and $\Oc>0$. The lower limit on the bulk cosmological constant is $-(1+\Oc)<\Olb$, which follows from the requirement that the square root in \eqref{omega-r2} should be real. Therefore, we impose the following prior on the bulk cosmological constant: 
\beq
-(1+\Oc)<\Olb \leq 1.0 \, .
\eeq

\subsection{Results}
We test the three models, described above, against the distances inferred from the SNe, BAO and CMB observations. We employ the \lstinline{emcee} \citep{emcee} package to fit the models using Markov chain Monte Carlo sampling.

\subsubsection{Analysing low-$z$ and high-$z$ BAO data separately, without considering reduced CMB parameters}
In the literature, the constrains on parameters from the high redshift BAO measurements, based on \lya~ and quasars, have been found to be slightly inconsistent with those from the low redshift galaxy BAO data. This discrepancy has been observed in different models including both $\Lambda$CDM (see \citet{bao,Okamatsu:2021zao}) and phantom braneworld models (see \citet{late-ujjaini}). Here we investigate if this discrepancy still exists in the current BAO datasets, and in this phantom braneworld  framework. In doing so, we divide the BAO data listed in table \ref{tab:bao} into two parts, namely, low-redshift ($z<1$) and high-redshift ($z>1$) datasets, and use them separately (and in conjunction) in the fits. To clearly see the inconsistency, if any, we do not additionally include the reduced CMB data here as they pose stringent constraints. Rather, we consider the {\em big-bang nucleosynthesis} (BBN) constraint \citep{Cooke:2017cwo} together with the JLA compilation of SNe data.

\begin{figure}
\centering
\subfigure[$\Lambda$CDM]{
\includegraphics[scale=0.85]{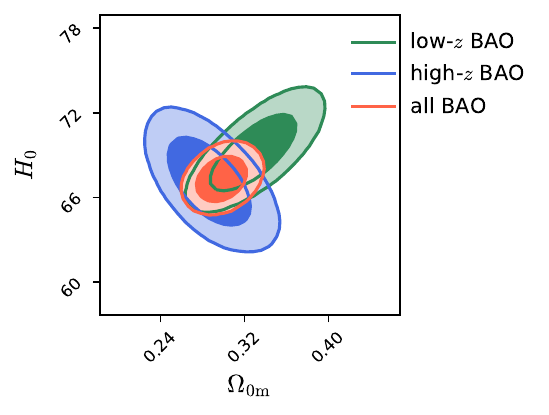}\label{fig:lcdm_bao}}\hspace*{-10 mm}
\subfigure[minimal phantom brane]{
\includegraphics[scale=0.85]{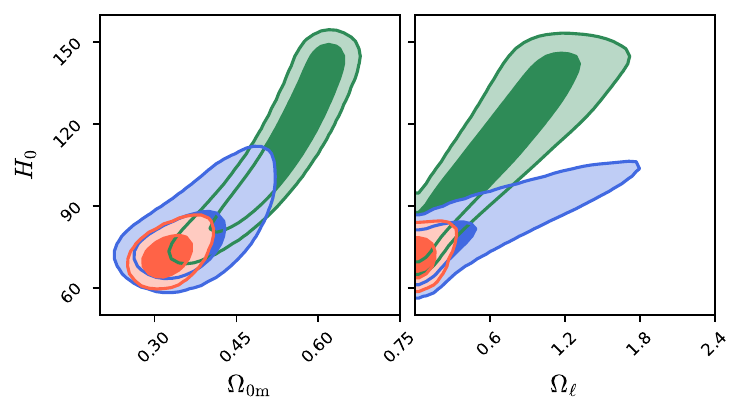}\label{fig:phbw_bao}}
\caption{The $68\%$ and $95\%$ confidence levels on the cosmological parameters are shown separately considering the low-$z$ BAO data (green), the high-$z$ BAO data (blue) and all the BAO data (red), in conjunction with SNe+BBN prior. The left panel shows the results for the $\Lambda$CDM model whereas the results for the minimal phantom brane are shown in the two figures of the right panel.}
\label{fig:bao_cls}
\end{figure}

Figure \ref{fig:lcdm_bao} shows the $68\%$ and $95\%$ confidence levels in the $\omt$$-$$H_0$ parameter space in the $\Lambda$CDM model. The green and blue contours are obtained using low and high redshift BAO data respectively while the red contour represents the results for the full BAO data. It is evident that low-$z$ and high-$z$ BAO datasets  impose constraints  on the $\omt$$-$$H_0$ plane in different directions. For low-$z$ and high-$z$ BAO data, a higher $H_0$ leads to higher  and lower values of $\omt$, respectively; however, the confidence levels are consistent within $1\sigma$.
Furthermore, the galaxy BAO data prefers a higher value of $H_0$ as compared to the high-$z$ BAO data. However, when considering all the BAO measurements together, we found that the parameter space is tightly constrained, as evident from the red contour. 
The discrepancy between the constraints from the low-$z$ and high-$z$ BAO data is much less significant in the newer BOSS DR16 data as compared to that in the earlier data sets (compare with figure 4 of \citet{bao} that assumes SDSS DR11 data, see also \citet{Okamatsu:2021zao}).

Figure \ref{fig:phbw_bao} illustrates the results for the minimal phantom brane model. The green, blue and red contours represent the $68\%$ and $95\%$ confidence levels (on the parameters $\omt$, $H_0$ and $\oml$) for low-$z$, high-$z$ and all BAO measurements respectively. The confidence levels on $\omt$ and $H_0$ are much more extended as compared to those of $\Lambda$CDM model. Once again we find that the high-$z$ \lya-QSO BAO data prefer a lower value of $H_0$ than the galaxy BAO data. Interestingly, the constraints for low-$z$ and high-$z$ BAO data are in the same direction, a higher $H_0$ leads to higher $\omt$ values (very high values of $\omt$ can be ruled out by other observations).  Remarkably, the galaxy and \lya-QSO BAO data separately allow large values of the brane parameter, $\oml \lesssim 1.8$ at $2\sigma$ for both. However, when we consider the full BAO dataset, the constraints become much tighter, $\oml \lesssim 0.3$ at $2\sigma$.
Comparing with figure 6 of \citet{late-ujjaini} (that uses BOSS DR12 BAO data), we find that the newer BAO dataset diminishes the inconsistency between the results from low-$z$ and high-$z$ BAO data.

\subsubsection{Considering reduced CMB data along with all BAO measurements}
\begin{table}
\centering
\renewcommand{\arraystretch}{1.5}
\setlength{\tabcolsep}{8pt}
\begin{tabular}{|c|c|c|c|c|c|}
    \hline
    
    Model name & $\omt$ & \makecell{$H_0$ \\ (km s$^{-1}$ Mpc$^{-1}$)}& $\omb$ & best-fit $\chi^2$ & $\Delta \chi ^2$  \\
    \hline
    \multicolumn{6}{|c|}{JLA + BAO + CMB}\\
    \hline
    \makecell{$\Lambda$CDM\\ $\omt$, $H_0$, $\omb$ } & $  {0.3131} ^{+0.0058}_{-0.0057}$ & $ {67.60} ^{+0.42}_{-0.42} $ & $ {0.0492} ^{+0.0005}_{-0.0005}$  &  $  697.76 $ & $  0$  \\
    \hline

    \makecell{Minimal phantom brane \\ $\omt$, $H_0$, $\omb$, $\oml$ } 
& $ {0.3051} ^{+0.0065}_{-0.0066} $ & $ {68.85} ^{+0.75}_{-0.68} $ & $ {0.0472} ^{+0.0010}_{-0.0011} $  &  $ 696.85 $ & $ -0.91 $  \\
    \hline
    
    \makecell{General phantom brane \\
    $\omt$, $H_0$, $\omb$, $\oml$, $\Oc$, $\Olb$} & ${0.3037} ^{+0.0065}_{-0.0065} $  & $ {69.08} ^{+0.71}_{-0.70} $ & $ {0.0468} ^{+0.0010}_{-0.0010}$ &  $ 696.86 $ & $ -0.90 $  \\
    \hline
    \hline
    \multicolumn{6}{|c|}{JLA + BAO + CMB + R19}\\
    \hline
    \makecell{$\Lambda$CDM \\ $\omt$, $H_0$, $\omb$ } & $ {0.3060} ^{+0.0055}_{-0.0054} $ & $ {68.13} ^{+0.41}_{-0.41} $ & $ {0.0486} ^{+0.0005}_{-0.0005}$  &  $  716.51 $ & $  0$  \\
    \hline

    \makecell{Minimal phantom brane \\ $\omt$, $H_0$, $\omb$, $\oml$ }  & $ {0.2956} ^{+0.0061}_{-0.006} $ & $ {70.01} ^{+0.71}_{-0.69} $ & $ {0.0456} ^{+0.0010}_{-0.0010} $ &  $ 708.67 $ & $ -7.84 $  \\
    \hline
    
    \makecell{General phantom brane \\
    $\omt$, $H_0$, $\omb$, $\oml$, $\Oc$, $\Olb$} & ${0.2945} ^{+0.006}_{-0.006} $  & $ {70.21} ^{+0.71}_{-0.70} $ & $ {0.0453} ^{+0.0010}_{-0.0010}$ &  $ 708.19 $ & $ -8.32 $  \\
    \hline

\end{tabular}
\caption{The top part of the table shows the median and $68\%$ percentile of the cosmological parameters in the three models considering the Type-1a SNe, BAO and CMB observations (without including the R19 measurement of $H_0$ value). The lower part shows the same but for considering the R19 prior along with the other observations. 
Here, $\Delta \chi^2$ is the difference in best-fit values of $\chi^2$ between a given model and $\Lambda$CDM model.
}
\label{table:summary}
\end{table}

\begin{figure}
\begin{center}
\includegraphics[width=0.5\textwidth]{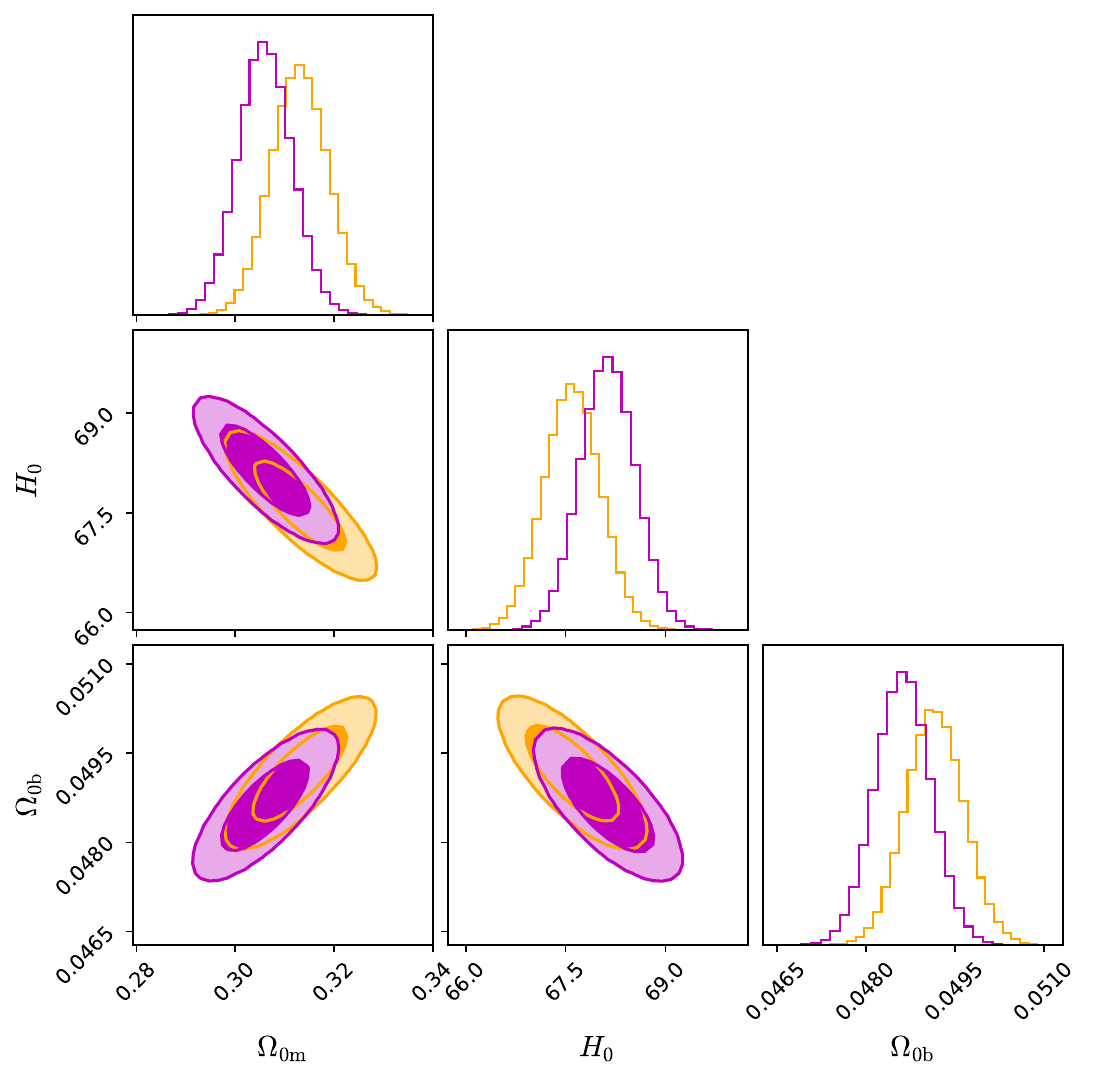}
\caption{$68\%$ and $95\%$ confidence levels on parameters for the $\Lambda$CDM model are illustrated here. The orange and magenta contours portray the results excluding and including the R19 prior on $H_0$.}
\label{fig:lcdm_cls}
\end{center}
\end{figure}

\begin{figure}
\begin{center}
\includegraphics[width=0.65\textwidth]{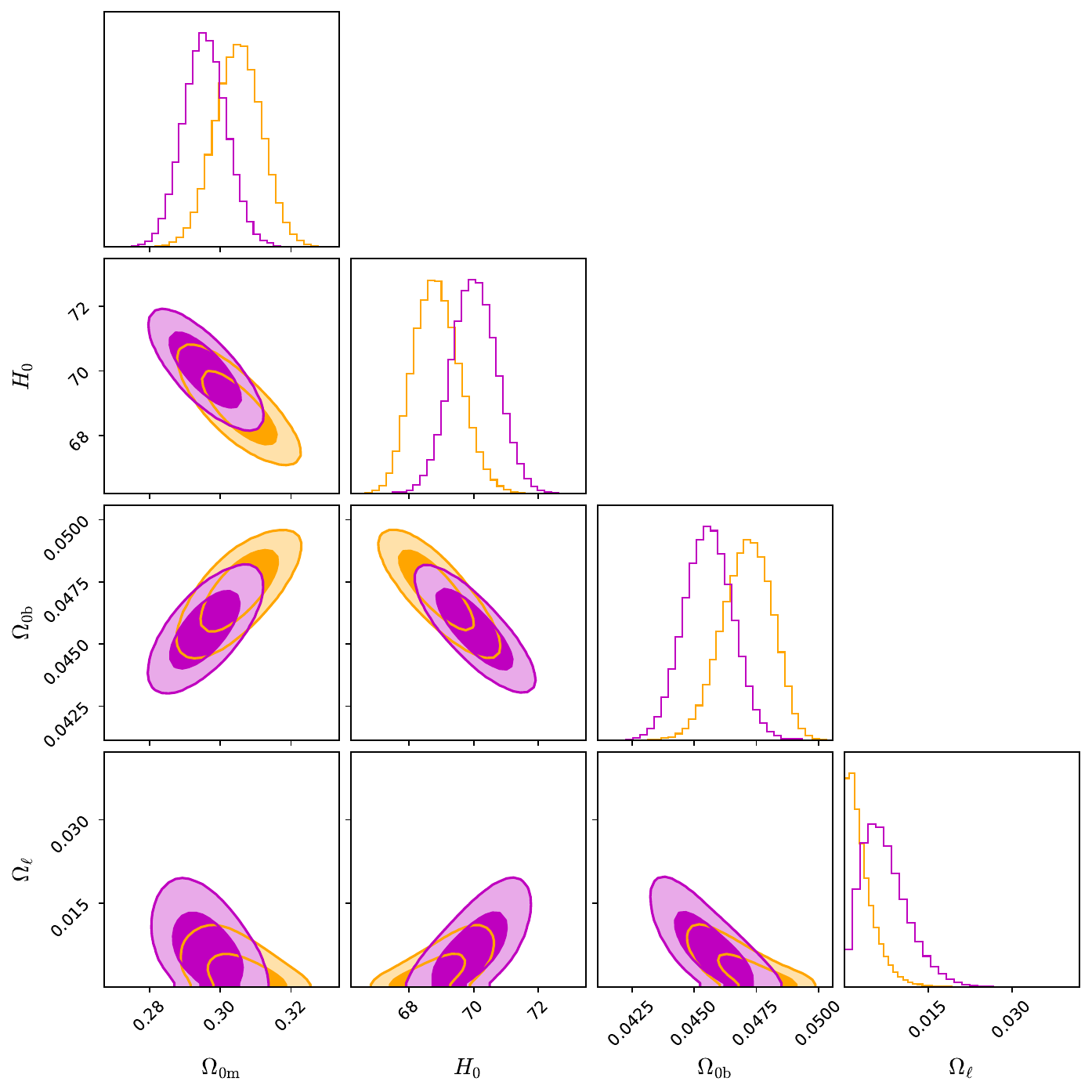}
\caption{$68\%$ and $95\%$ confidence levels on the parameters of the minimal phantom brane model are shown here. The orange and magenta contours portray the results excluding and including the R19 prior on $H_0$, respectively. Notice that upon considering the R19 data point, the fit prefers a nonzero $\oml \sim 0.005$ which implies that the presence of an extra dimension is favoured by the data.}
\label{fig:phbwcls}
\end{center}
\end{figure}

Now we consider the reduced CMB data together with all the measurements of BAO and SNe.
The $68\%$ and $95\%$ confidence levels (CL) of the cosmological parameters for $\Lambda$CDM and minimal phantom brane model are shown in figures \ref{fig:lcdm_cls} and \ref{fig:phbwcls}, respectively. The orange contours in both figures represent the CL's without including the local measurement of $H_0$ by \citet{riess19} (R19), whereas the CL's considering the R19 prior are shown by the magenta contours. 

Without considering R19, we get $H_0=67.60\pm 0.42$ (the median and $68 \%$ around it) from the fit for the $\Lambda$CDM model. This is consistent with the {\em Planck\/} 2018 analysis with the full CMB angular power spectrum data \citep{planck2018}. This preferred value of $H_0$ in $\Lambda$CDM model lies significantly below the R19 measurement, $H_0=74.03\pm1.42$, amounting to the celebrated $\sim 4.4 \sigma$ tension.
When we include the R19 measurement as a prior, the $H_0$ posterior in the $\Lambda$CDM model only shifts slightly, as illustrated by the magenta contours in figure \ref{fig:lcdm_cls}, leading to a rise in best-fit $\chi^2$ value. This further demonstrates the inconsistency between the $\Lambda$CDM model predicted by {\em Planck\/}
and the R19 measurement. 

On the other hand, the minimal phantom brane intrinsically prefers a higher $H_0$ value, as illustrated in figure \ref{fig:phbwcls}, because of the slower expansion rate in comparison with the $\Lambda$CDM model. Without considering the R19 prior, the posterior is given by $H_0={68.85}^{+0.75}_{-0.68}$ (the median and the $68\%$ percentile around it). Note that not only does the uncertainty in $H_0$ increase slightly as compared to that in $\Lambda$CDM, but the mean/median of $H_0$ also shifts to a larger value, which results in a much lower ($\sim 3.2 \sigma$) tension with R19 measurement. Therefore, the minimal phantom brane model cannot completely resolve the so-called $H_0$ tension considering the present datasets but reduces the tension significantly. The brane parameter is quite tightly constrained in this case, $\oml \lesssim 0.01$ at $2\sigma$ which is slightly narrower than the previous findings for the same model but with earlier datasets; see, e.g., \citet{late-ujjaini, Lazkoz:2006gp}.

When we include the R19 measurement in the analysis, the $H_0$ confidence levels are further pushed to higher values, $H_0= {70.01} ^{+0.71}_{-0.69}$. Now the fit prefers a non-zero brane parameter $\oml \approx 0.005$, which supports the existence of the extra dimension. The confidence levels for it also extend significantly, $\oml \lesssim 0.02$ at $2\sigma$.
The best-fit $\chi^2$ values in this case is $7.84$ lower than that in $\Lambda$CDM model. This indicates that the minimal phantom brane model is much more consistent with the high $H_0$ values than the $\Lambda$CDM model.

A summary of the results, precisely the median, along with the $68\%$ percentile around it, of the cosmological parameters in the different models are given in table \ref{table:summary}. The top and bottom parts present the results excluding and including the R19 prior, respectively. The last two columns respectively show the best-fit $\chi^2$ values in the fits and the  differences in the $\chi^2$ value with respect to the $\Lambda$CDM model.

Even if we include the bulk cosmological constant ($\Olb$) together with dark radiation ($\Oc$) on the phantom brane, the fit for the cosmological parameters are quite similar to that of the minimal phantom brane, in which these two parameters are equal to zero. The $H_0$ confidence levels are pushed  towards higher value, but only slightly (the value of $\omb$ decreases accordingly). We do not find any support for the cosmic loitering feature \citep{loiter} that requires the dark radiation term dominating over other terms under the square-root in \eqref{hubble_loiter}. 

\begin{figure}
\begin{center}
\includegraphics[width=0.49\textwidth]{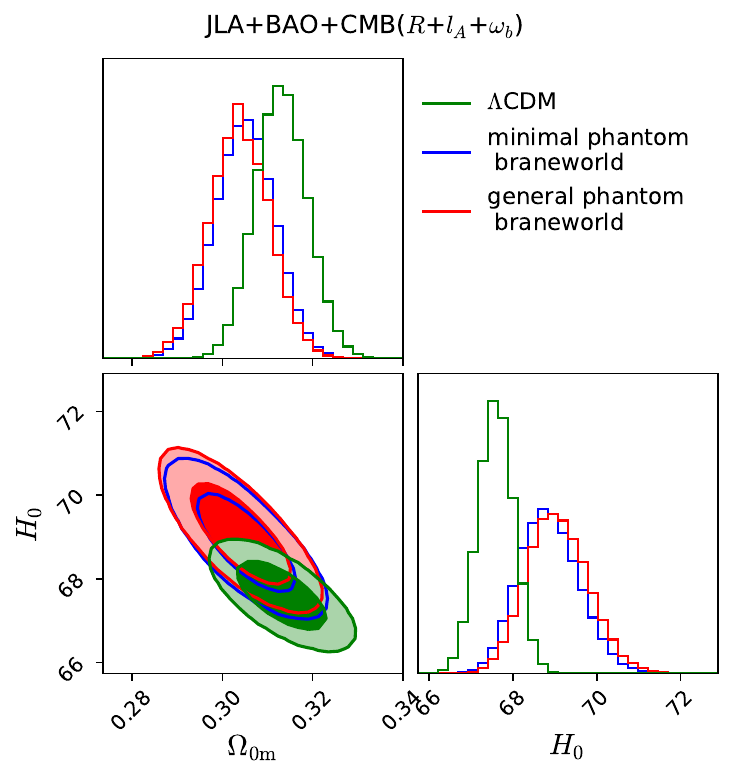}
\includegraphics[width=0.49\textwidth]{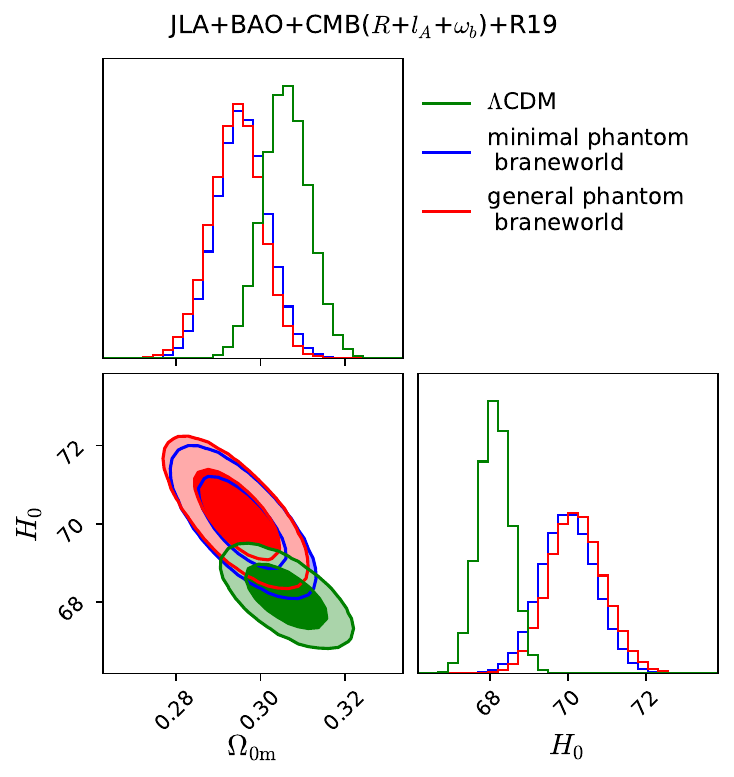}
\caption{$68\%$ and $95\%$ confidence levels (CL) on the $\omt$$-$$H_0$ plane are compared for the $\Lambda$CDM (green), minimal phantom brane (blue) and general phantom brane (red) models. The right panel includes the R19 prior (the left panel does not). The CL's for both phantom brane models are shifted towards slightly higher $H_0$ (and slightly lower $\omt$) values as compared to the $\Lambda$CDM model. This shift is larger when we include the R19 prior as illustrated by the right panel.}
\label{fig:comp}
\end{center}
\end{figure}

The confidence levels on the $\omt$$-$$H_0$ plane for all the three models ($\Lambda$CDM, minimal and general phantom brane) have been compared in figure \ref{fig:comp}. The left and right panels portray the results of excluding and including the R19 prior, respectively. It is clearly evident from the left panel that even without R19 prior larger values of $H_0$ are preferred in the braneworld models as compared to \lcdm~ model.
Comparing the left and right panels, one notices that inclusion of the R19 pushes all the confidence levels towards higher $H_0$ values, the shifts are larger for the braneworld models.
From both panels it is also evident that inclusion of the bulk cosmological constant and dark radiation on the  phantom brane does not change the $\omt$$-$$H_0$ confidence levels, aside from producing a small increase in $H_0$.

\section{Conclusions and discussion}
\label{sec:con}

Braneworld models are string-theory-inspired phenomenological models with large (often noncompact) extra dimensions. They were motivated by earlier studies of relations between string theory and supergravity \citep{Horava:1995qa}, and later used as theoretical schemes for addressing the Planck hierarchy problem \citep{Antoniadis:1998ig, Arkani-Hamed:1998jmv, Randall:1999ee, Randall:1999vf}.  Their subsequent application to cosmology  opened up a new field of research in this interdisciplinary region.

One popular version that attracted much attention in cosmology is the braneworld model with a single large extra dimension and with `induced gravity' on the brane \citep{Dvali:2000hr, Collins:2000yb, Shtanov:2000vr}.  Among other features, quite incidentally, it was discovered that the stable branch of solutions of this model describes effective dark energy with supernegative equation of state \citep{ss02, Lue:2004za}. Because of this phantom-like behaviour (hence the term `phantom brane' \citep{Bag:2016tvc}), one can also expect that cosmological tests of this model would yield higher values of $H_0$. This observation lies behind the idea of alleviating the $H_0$ tension in cosmological scenarios based on this model.
Note that the braneworld cosmological model under discussion does not suffer from the usual instabilities associated with a phenomenological phantom field; it rather smoothly passes on to a De~Sitter phase in the future without running into a `big-rip' singularity.

In this work, we test the background expansion on the phantom brane against the observations of SNe, BAO and CMB (expressed in terms of background parameters). Compared to the $\Lambda$CDM model, the minimal phantom brane model contains one additional free parameter, the brane parameter $\Omega_\ell$, which goes to zero as the model smoothly passes to $\Lambda$CDM (see section \ref{sec:minimal}). Before including the compressed CMB data, we first test the internal consistency between the results from the low-redshift galaxy BAO data and the high-redshift \lya-QSO BAO measurements. For low-$z$ and high-$z$ BAO data, used separately in conjunction with SNe data (and the BBN prior), the constraints on the parameters are consistent within $1\sigma$ in the phantom braneworld scenario. The directions of degeneracy in the $\omt$$-$$H_0$ plane are also found to be consistent for these two BAO datasets on the minimal phantom brane (in contrast to $\Lambda$CDM); higher value of $H_0$ leads to larger values of $\omt$. However, low-$z$ BAO data allow for significantly higher values of $H_0$ than the high-$z$ \lya-QSO BAO data. Both datasets individually allow for large values of the brane parameter, $\oml \lesssim 1.8$ at $2\sigma$. When we include all the BAO measurements into the analysis, the parameter space is tightly constrained, $\oml \lesssim 0.3$ at $2\sigma$.

Upon inclusion, the compressed CMB data constrain the parameter space very tightly.
Using SNe+BAO+CMB datasets, we observe that the minimal phantom brane model prefers a significantly higher value of $H_0$ as compared to the $\Lambda$CDM model even disregarding the R19 local measurement of $H_0$. The constraint on the brane parameter, $\oml \lesssim 0.01$ at $2\sigma$, appears to be tighter as compared to the previous analyses with earlier datasets; e.g., compare with the results of \citet{late-ujjaini, Schmidt:2009sv,Lazkoz:2006gp}. 
Although the phantom brane model cannot completely resolve the $H_0$ tension, it reduces the tension significantly, from the current $ 4.4\,\sigma$ tension in the $\Lambda$CDM model to nearly $3.2\,\sigma$ tension in the minimal phantom brane model.

If we include the R19 local measurement as a prior on $H_0$, the fit prefers even higher values of $H_0$ (and lower values of $\omt$) in the minimal phantom brane model. The allowed range of the brane parameter is now significantly extended, $\oml \lesssim 0.02$ at $2\sigma$.
The fit also prefers a nonzero value of $\oml \approx 0.005$ (best-fit), favouring the existence of extra dimension. Moreover, the best-fit $\chi^2$ in the minimal phantom brane model is significantly lower than that in the $\Lambda$CDM model; $\Delta \chi^2=-7.84$. This indicates that the phantom brane model can accommodate a high value of $H_0$ more easily than $\Lambda$CDM. 

When we consider the general phantom brane model by including the bulk cosmological constant and dark radiation (which adds two new parameters to the model compared to its minimal version, see section \ref{sec:general}), we find that the fits for the cosmological parameters are quite similar to those for the minimal phantom brane. The preferred values of $H_0$ become slightly higher than those in the minimal phantom brane. This general spatially flat braneworld model can also imitate the evolution of a spatially closed $\Lambda$CDM when the dark-radiation term dominates at late times \citep{loiter}, but we do not find any strong evidence from the data in support of this scenario.

In conclusion, we should emphasise that the braneworld scenarios considered in this work naturally describe an effective dark energy that exhibits phantom-like behaviour at late times. This phantom-like behaviour can accommodate larger values of the Hubble constant, although the braneworld models at present cannot completely reconcile all the cosmological datasets including the local measurements of $H_0$. Other models of late-time evolution of dark energy also have difficulties in achieving this \citep{late-arman2, DiValentino:2019dzu, Arendse:2019hev} (a detailed discussion of modified gravity models in relation to the Hubble
tension is contained in section 10 of  \citet{DiValentino:2021izs}), which may be a general feature of the late-time physics approach to this problem \citep{Lemos:2018smw,Benevento:2020fev,Camarena:2021jlr,Efstathiou:2021ocp}. We should note that finding a significant systematic in any of the cosmological datasets can change the current situation dramatically, hence we should keep an open mind on different possible options \citep{Keeley:2020aym}. In this regard, we note that the braneworld scenario, being physically well motivated, remains to be a viable candidate describing the current cosmological observations.

\section*{Acknowledgements}
\label{sec:ack}
The supercomputing clusters, Nurion at the Korea Institute of Science and Technology Information (KISTI) and the Seondeok at KASI, have been used in this work.
S.~B. thanks Ryan Keeley for many useful discussions. 
V.~S. was partially supported by the J.C.~Bose fellowship of the Department of Science and Technology, Government of India.
The work of Y.~S. was supported by the National Research Foundation of Ukraine under Project No.~2020.02/0073.

\bibliographystyle{aasjournal}
\bibliography{bw_test}

\begin{thebibliography}{}
\expandafter\ifx\csname natexlab\endcsname\relax\def\natexlab#1{#1}\fi
\providecommand{\url}[1]{\href{#1}{#1}}
\providecommand{\dodoi}[1]{doi:~\href{http://doi.org/#1}{\nolinkurl{#1}}}
\providecommand{\doeprint}[1]{\href{http://ascl.net/#1}{\nolinkurl{http://ascl.net/#1}}}
\providecommand{\doarXiv}[1]{\href{https://arxiv.org/abs/#1}{\nolinkurl{https://arxiv.org/abs/#1}}}

\bibitem[{{Agrawal} {et~al.}(2019){Agrawal}, {Cyr-Racine}, {Pinner}, \&
  {Randall}}]{early-randall}
{Agrawal}, P., {Cyr-Racine}, F.-Y., {Pinner}, D., \& {Randall}, L. 2019, arXiv
  e-prints, arXiv:1904.01016.
\newblock \doarXiv{1904.01016}

\bibitem[{{Alam} {et~al.}(2017){Alam}, {Ata}, {Bailey}, {Beutler}, {Bizyaev},
  {Blazek}, {Bolton}, {Brownstein}, {Burden}, {Chuang}, {Comparat}, {Cuesta},
  {Dawson}, {Eisenstein}, {Escoffier}, {Gil-Mar{\'\i}n}, {Grieb}, {Hand}, {Ho},
  {Kinemuchi}, {Kirkby}, {Kitaura}, {Malanushenko}, {Malanushenko}, {Maraston},
  {McBride}, {Nichol}, {Olmstead}, {Oravetz}, {Padmanabhan},
  {Palanque-Delabrouille}, {Pan}, {Pellejero-Ibanez}, {Percival}, {Petitjean},
  {Prada}, {Price-Whelan}, {Reid}, {Rodr{\'\i}guez-Torres}, {Roe}, {Ross},
  {Ross}, {Rossi}, {Rubi{\~n}o-Mart{\'\i}n}, {Saito}, {Salazar-Albornoz},
  {Samushia}, {S{\'a}nchez}, {Satpathy}, {Schlegel}, {Schneider},
  {Sc{\'o}ccola}, {Seo}, {Sheldon}, {Simmons}, {Slosar}, {Strauss}, {Swanson},
  {Thomas}, {Tinker}, {Tojeiro}, {Maga{\~n}a}, {Vazquez}, {Verde}, {Wake},
  {Wang}, {Weinberg}, {White}, {Wood-Vasey}, {Y{\`e}che}, {Zehavi}, {Zhai}, \&
  {Zhao}}]{sdss12}
{Alam}, S., {Ata}, M., {Bailey}, S., {et~al.} 2017, \mnras, 470, 2617,
  \dodoi{10.1093/mnras/stx721}

\bibitem[{Alam {et~al.}(2021)}]{eBOSS:2020yzd}
Alam, S., {et~al.} 2021, Phys. Rev. D, 103, 083533,
  \dodoi{10.1103/PhysRevD.103.083533}

\bibitem[{Alam {et~al.}(2017)Alam, Bag, \& Sahni}]{late-ujjaini}
Alam, U., Bag, S., \& Sahni, V. 2017, Phys. Rev. D, 95, 023524,
  \dodoi{10.1103/PhysRevD.95.023524}

\bibitem[{Antoniadis {et~al.}(1998)Antoniadis, Arkani-Hamed, Dimopoulos, \&
  Dvali}]{Antoniadis:1998ig}
Antoniadis, I., Arkani-Hamed, N., Dimopoulos, S., \& Dvali, G.~R. 1998, Phys.
  Lett. B, 436, 257, \dodoi{10.1016/S0370-2693(98)00860-0}

\bibitem[{{Arendse} {et~al.}(2020){Arendse}, {Wojtak}, {Agnello}, {Chen},
  {Fassnacht}, {Sluse}, {Hilbert}, {Millon}, {Bonvin}, {Wong}, {Courbin},
  {Suyu}, {Birrer}, {Treu}, \& {Koopmans}}]{Arendse:2019hev}
{Arendse}, N., {Wojtak}, R.~J., {Agnello}, A., {et~al.} 2020, Astron.
  Astrophys., 639, A57, \dodoi{10.1051/0004-6361/201936720}

\bibitem[{Arkani-Hamed {et~al.}(1998)Arkani-Hamed, Dimopoulos, \&
  Dvali}]{Arkani-Hamed:1998jmv}
Arkani-Hamed, N., Dimopoulos, S., \& Dvali, G.~R. 1998, Phys. Lett. B, 429,
  263, \dodoi{10.1016/S0370-2693(98)00466-3}

\bibitem[{{Aubourg} {et~al.}(2015){Aubourg}, {Bailey}, {Bautista}, {Beutler},
  {Bhardwaj}, {Bizyaev}, {Blanton}, {Blomqvist}, {Bolton}, {Bovy},
  {Brewington}, {Brinkmann}, {Brownstein}, {Burden}, {Busca}, {Carithers},
  {Chuang}, {Comparat}, {Croft}, {Cuesta}, {Dawson}, {Delubac}, {Eisenstein},
  {Font-Ribera}, {Ge}, {Le Goff}, {Gontcho}, {Gott}, {Gunn}, {Guo}, {Guy},
  {Hamilton}, {Ho}, {Honscheid}, {Howlett}, {Kirkby}, {Kitaura}, {Kneib},
  {Lee}, {Long}, {Lupton}, {Maga{\~n}a}, {Malanushenko}, {Malanushenko},
  {Manera}, {Maraston}, {Margala}, {McBride}, {Miralda-Escud{\'e}}, {Myers},
  {Nichol}, {Noterdaeme}, {Nuza}, {Olmstead}, {Oravetz}, {P{\^a}ris},
  {Padmanabhan}, {Palanque-Delabrouille}, {Pan}, {Pellejero-Ibanez},
  {Percival}, {Petitjean}, {Pieri}, {Prada}, {Reid}, {Rich}, {Roe}, {Ross},
  {Ross}, {Rossi}, {Rubi{\~n}o-Mart{\'\i}n}, {S{\'a}nchez}, {Samushia},
  {G{\'e}nova-Santos}, {Sc{\'o}ccola}, {Schlegel}, {Schneider}, {Seo},
  {Sheldon}, {Simmons}, {Skibba}, {Slosar}, {Strauss}, {Thomas}, {Tinker},
  {Tojeiro}, {Vazquez}, {Viel}, {Wake}, {Weaver}, {Weinberg}, {Wood-Vasey},
  {Y{\`e}che}, {Zehavi}, {Zhao}, \& {BOSS Collaboration}}]{bao}
{Aubourg}, {\'E}., {Bailey}, S., {Bautista}, J.~E., {et~al.} 2015, Phys. Rev.
  D, 92, 123516, \dodoi{10.1103/PhysRevD.92.123516}

\bibitem[{Bag {et~al.}(2018)Bag, Mishra, \& Sahni}]{Bag:2018jle}
Bag, S., Mishra, S.~S., \& Sahni, V. 2018, Phys. Rev. D, 97, 123537,
  \dodoi{10.1103/PhysRevD.97.123537}

\bibitem[{Bag {et~al.}(2016)Bag, Viznyuk, Shtanov, \& Sahni}]{Bag:2016tvc}
Bag, S., Viznyuk, A., Shtanov, Y., \& Sahni, V. 2016, JCAP, 07, 038,
  \dodoi{10.1088/1475-7516/2016/07/038}

\bibitem[{Benetti {et~al.}(2019)Benetti, Miranda, Borges, Pigozzo, Carneiro, \&
  Alcaniz}]{late-jailson}
Benetti, M., Miranda, W., Borges, H.~A., {et~al.} 2019, JCAP, 12, 023,
  \dodoi{10.1088/1475-7516/2019/12/023}

\bibitem[{Benevento {et~al.}(2020)Benevento, Hu, \& Raveri}]{Benevento:2020fev}
Benevento, G., Hu, W., \& Raveri, M. 2020, Phys. Rev. D, 101, 103517,
  \dodoi{10.1103/PhysRevD.101.103517}

\bibitem[{{Betoule} {et~al.}(2014){Betoule}, {Kessler}, {Guy}, {Mosher},
  {Hardin}, {Biswas}, {Astier}, {El-Hage}, {Konig}, {Kuhlmann}, {Marriner},
  {Pain}, {Regnault}, {Balland}, {Bassett}, {Brown}, {Campbell}, {Carlberg},
  {Cellier-Holzem}, {Cinabro}, {Conley}, {D'Andrea}, {DePoy}, {Doi}, {Ellis},
  {Fabbro}, {Filippenko}, {Foley}, {Frieman}, {Fouchez}, {Galbany}, {Goobar},
  {Gupta}, {Hill}, {Hlozek}, {Hogan}, {Hook}, {Howell}, {Jha}, {Le Guillou},
  {Leloudas}, {Lidman}, {Marshall}, {M{\"o}ller}, {Mour{\~a}o}, {Neveu},
  {Nichol}, {Olmstead}, {Palanque-Delabrouille}, {Perlmutter}, {Prieto},
  {Pritchet}, {Richmond}, {Riess}, {Ruhlmann-Kleider}, {Sako}, {Schahmaneche},
  {Schneider}, {Smith}, {Sollerman}, {Sullivan}, {Walton}, \&
  {Wheeler}}]{Betoule:2014frx}
{Betoule}, M., {Kessler}, R., {Guy}, J., {et~al.} 2014, Astron. Astrophys.,
  568, A22, \dodoi{10.1051/0004-6361/201423413}

\bibitem[{{Beutler} {et~al.}(2011){Beutler}, {Blake}, {Colless}, {Jones},
  {Staveley-Smith}, {Campbell}, {Parker}, {Saunders}, \& {Watson}}]{6df}
{Beutler}, F., {Blake}, C., {Colless}, M., {et~al.} 2011, \mnras, 416, 3017,
  \dodoi{10.1111/j.1365-2966.2011.19250.x}

\bibitem[{{Birrer} {et~al.}(2019){Birrer}, {Treu}, {Rusu}, {Bonvin},
  {Fassnacht}, {Chan}, {Agnello}, {Shajib}, {Chen}, {Auger}, {Courbin},
  {Hilbert}, {Sluse}, {Suyu}, {Wong}, {Marshall}, {Lemaux}, \&
  {Meylan}}]{birrer}
{Birrer}, S., {Treu}, T., {Rusu}, C.~E., {et~al.} 2019, Mon. Not. Roy. Astron.
  Soc., 484, 4726, \dodoi{10.1093/mnras/stz200}

\bibitem[{{Bonvin} {et~al.}(2017){Bonvin}, {Courbin}, {Suyu}, {Marshall},
  {Rusu}, {Sluse}, {Tewes}, {Wong}, {Collett}, {Fassnacht}, {Treu}, {Auger},
  {Hilbert}, {Koopmans}, {Meylan}, {Rumbaugh}, {Sonnenfeld}, \&
  {Spiniello}}]{bonvin}
{Bonvin}, V., {Courbin}, F., {Suyu}, S.~H., {et~al.} 2017, \mnras, 465, 4914,
  \dodoi{10.1093/mnras/stw3006}

\bibitem[{{Breuval} {et~al.}(2019){Breuval}, {Kervella}, {Arenou}, {Bono},
  {Gallenne}, {Trahin}, {M{\'e}rand}, {Storm}, {Inno}, {Pietrzynski}, {Gieren},
  {Nardetto}, {Graczyk}, {Borgniet}, {Javanmardi}, \& {Hocd{\'e}}}]{breuval19}
{Breuval}, L., {Kervella}, P., {Arenou}, F., {et~al.} 2019, arXiv e-prints,
  arXiv:1910.04694.
\newblock \doarXiv{1910.04694}

\bibitem[{Camarena \& Marra(2021)}]{Camarena:2021jlr}
Camarena, D., \& Marra, V. 2021, Mon. Not. Roy. Astron. Soc., 504, 5164,
  \dodoi{10.1093/mnras/stab1200}

\bibitem[{Charmousis {et~al.}(2006)Charmousis, Gregory, Kaloper, \&
  Padilla}]{Charmousis:2006pn}
Charmousis, C., Gregory, R., Kaloper, N., \& Padilla, A. 2006, JHEP, 10, 066,
  \dodoi{10.1088/1126-6708/2006/10/066}

\bibitem[{Chen {et~al.}(2019)Chen, Huang, \& Wang}]{Chen:2018dbv}
Chen, L., Huang, Q.-G., \& Wang, K. 2019, JCAP, 02, 028,
  \dodoi{10.1088/1475-7516/2019/02/028}

\bibitem[{Collins \& Holdom(2000)}]{Collins:2000yb}
Collins, H., \& Holdom, B. 2000, Phys. Rev. D, 62, 105009,
  \dodoi{10.1103/PhysRevD.62.105009}

\bibitem[{Cooke {et~al.}(2018)Cooke, Pettini, \& Steidel}]{Cooke:2017cwo}
Cooke, R.~J., Pettini, M., \& Steidel, C.~C. 2018, Astrophys. J., 855, 102,
  \dodoi{10.3847/1538-4357/aaab53}

\bibitem[{Dainotti {et~al.}(2021)Dainotti, De~Simone, Schiavone, Montani,
  Rinaldi, \& Lambiase}]{Dainotti:2021pqg}
Dainotti, M.~G., De~Simone, B., Schiavone, T., {et~al.} 2021, Astrophys. J.,
  912, 150, \dodoi{10.3847/1538-4357/abeb73}

\bibitem[{Deffayet(2001)}]{Deffayet:2000uy}
Deffayet, C. 2001, Phys. Lett. B, 502, 199,
  \dodoi{10.1016/S0370-2693(01)00160-5}

\bibitem[{Di~Valentino {et~al.}(2017)Di~Valentino, Melchiorri, \&
  Mena}]{late-melchiorri1}
Di~Valentino, E., Melchiorri, A., \& Mena, O. 2017, Phys. Rev. D, 96, 043503,
  \dodoi{10.1103/PhysRevD.96.043503}

\bibitem[{Di~Valentino {et~al.}(2020{\natexlab{a}})Di~Valentino, Melchiorri,
  Mena, \& Vagnozzi}]{late-valentino}
Di~Valentino, E., Melchiorri, A., Mena, O., \& Vagnozzi, S. 2020{\natexlab{a}},
  Phys. Dark Univ., 30, 100666, \dodoi{10.1016/j.dark.2020.100666}

\bibitem[{Di~Valentino {et~al.}(2016)Di~Valentino, Melchiorri, \&
  Silk}]{late-silk}
Di~Valentino, E., Melchiorri, A., \& Silk, J. 2016, Phys. Lett. B, 761, 242,
  \dodoi{10.1016/j.physletb.2016.08.043}

\bibitem[{Di~Valentino {et~al.}(2020{\natexlab{b}})Di~Valentino, Melchiorri, \&
  Silk}]{DiValentino:2019dzu}
---. 2020{\natexlab{b}}, JCAP, 01, 013, \dodoi{10.1088/1475-7516/2020/01/013}

\bibitem[{Di~Valentino {et~al.}(2021)Di~Valentino, Mena, Pan, Visinelli, Yang,
  Melchiorri, Mota, Riess, \& Silk}]{DiValentino:2021izs}
Di~Valentino, E., Mena, O., Pan, S., {et~al.} 2021, Class. Quant. Grav., 38,
  153001, \dodoi{10.1088/1361-6382/ac086d}

\bibitem[{Dvali {et~al.}(2000)Dvali, Gabadadze, \& Porrati}]{Dvali:2000hr}
Dvali, G.~R., Gabadadze, G., \& Porrati, M. 2000, Phys. Lett. B, 485, 208,
  \dodoi{10.1016/S0370-2693(00)00669-9}

\bibitem[{{Efstathiou}(2021)}]{Efstathiou:2021ocp}
{Efstathiou}, G. 2021, \mnras, 505, 3866, \dodoi{10.1093/mnras/stab1588}

\bibitem[{Evslin {et~al.}(2018)Evslin, Sen, \& Ruchika}]{early-anjan18}
Evslin, J., Sen, A.~A., \& Ruchika. 2018, Phys. Rev. D, 97, 103511,
  \dodoi{10.1103/PhysRevD.97.103511}

\bibitem[{{Fiorini} {et~al.}(2021){Fiorini}, {Koyama}, {Izard}, {Winther},
  {Wright}, \& {Li}}]{Fiorini:2021dzs}
{Fiorini}, B., {Koyama}, K., {Izard}, A., {et~al.} 2021, \jcap, 2021, 021,
  \dodoi{10.1088/1475-7516/2021/09/021}

\bibitem[{{Foreman-Mackey} {et~al.}(2013){Foreman-Mackey}, {Hogg}, {Lang}, \&
  {Goodman}}]{emcee}
{Foreman-Mackey}, D., {Hogg}, D.~W., {Lang}, D., \& {Goodman}, J. 2013, \pasp,
  125, 306, \dodoi{10.1086/670067}

\bibitem[{Ghosh {et~al.}(2020)Ghosh, Khatri, \& Roy}]{early-rishi}
Ghosh, S., Khatri, R., \& Roy, T.~S. 2020, Phys. Rev. D, 102, 123544,
  \dodoi{10.1103/PhysRevD.102.123544}

\bibitem[{{Gogoi} {et~al.}(2021){Gogoi}, {Kumar Sharma}, {Chanda}, \&
  {Das}}]{Gogoi:2020qif}
{Gogoi}, A., {Kumar Sharma}, R., {Chanda}, P., \& {Das}, S. 2021, \apj, 915,
  132, \dodoi{10.3847/1538-4357/abfe5b}

\bibitem[{Gorbunov {et~al.}(2006)Gorbunov, Koyama, \&
  Sibiryakov}]{Gorbunov:2005zk}
Gorbunov, D., Koyama, K., \& Sibiryakov, S. 2006, Phys. Rev. D, 73, 044016,
  \dodoi{10.1103/PhysRevD.73.044016}

\bibitem[{Hill {et~al.}(2020)Hill, McDonough, Toomey, \&
  Alexander}]{Hill:2020osr}
Hill, J.~C., McDonough, E., Toomey, M.~W., \& Alexander, S. 2020, Phys. Rev. D,
  102, 043507, \dodoi{10.1103/PhysRevD.102.043507}

\bibitem[{Hojjati {et~al.}(2013)Hojjati, Linder, \& Samsing}]{linder13}
Hojjati, A., Linder, E.~V., \& Samsing, J. 2013, Phys. Rev. Lett., 111, 041301,
  \dodoi{10.1103/PhysRevLett.111.041301}

\bibitem[{Ho\v{r}ava \& Witten(1996)}]{Horava:1995qa}
Ho\v{r}ava, P., \& Witten, E. 1996, Nucl. Phys. B, 460, 506,
  \dodoi{10.1016/0550-3213(95)00621-4}

\bibitem[{Howlett {et~al.}(2015)Howlett, Ross, Samushia, Percival, \&
  Manera}]{mgs2}
Howlett, C., Ross, A.~J., Samushia, L., Percival, W.~J., \& Manera, M. 2015,
  Mon. Not. Roy. Astron. Soc., 449, 848, \dodoi{10.1093/mnras/stu2693}

\bibitem[{Hu \& Sugiyama(1996)}]{Hu:1995en}
Hu, W., \& Sugiyama, N. 1996, Astrophys. J., 471, 542, \dodoi{10.1086/177989}

\bibitem[{Jedamzik \& Pogosian(2020)}]{Jedamzik:2020krr}
Jedamzik, K., \& Pogosian, L. 2020, Phys. Rev. Lett., 125, 181302,
  \dodoi{10.1103/PhysRevLett.125.181302}

\bibitem[{Joudaki {et~al.}(2018)Joudaki, Kaplinghat, Keeley, \&
  Kirkby}]{late-kaplinghat1}
Joudaki, S., Kaplinghat, M., Keeley, R., \& Kirkby, D. 2018, Phys. Rev. D, 97,
  123501, \dodoi{10.1103/PhysRevD.97.123501}

\bibitem[{Karwal \& Kamionkowski(2016)}]{early-kamion0}
Karwal, T., \& Kamionkowski, M. 2016, Phys. Rev. D, 94, 103523,
  \dodoi{10.1103/PhysRevD.94.103523}

\bibitem[{Keeley {et~al.}(2019)Keeley, Joudaki, Kaplinghat, \&
  Kirkby}]{late-kaplinghat2}
Keeley, R.~E., Joudaki, S., Kaplinghat, M., \& Kirkby, D. 2019, JCAP, 12, 035,
  \dodoi{10.1088/1475-7516/2019/12/035}

\bibitem[{Keeley {et~al.}(2021)Keeley, Shafieloo, Zhao, Vazquez, \&
  Koo}]{Keeley:2020aym}
Keeley, R.~E., Shafieloo, A., Zhao, G.-B., Vazquez, J.~A., \& Koo, H. 2021,
  Astron. J., 161, 151, \dodoi{10.3847/1538-3881/abdd2a}

\bibitem[{Knox \& Millea(2020)}]{knox19}
Knox, L., \& Millea, M. 2020, Phys. Rev. D, 101, 043533,
  \dodoi{10.1103/PhysRevD.101.043533}

\bibitem[{Koyama(2007)}]{Koyama:2007za}
Koyama, K. 2007, Class. Quant. Grav., 24, R231,
  \dodoi{10.1088/0264-9381/24/24/R01}

\bibitem[{Koyama \& Maartens(2006)}]{Koyama:2005kd}
Koyama, K., \& Maartens, R. 2006, JCAP, 01, 016,
  \dodoi{10.1088/1475-7516/2006/01/016}

\bibitem[{Lazkoz {et~al.}(2006)Lazkoz, Maartens, \& Majerotto}]{Lazkoz:2006gp}
Lazkoz, R., Maartens, R., \& Majerotto, E. 2006, Phys. Rev. D, 74, 083510,
  \dodoi{10.1103/PhysRevD.74.083510}

\bibitem[{Lemos {et~al.}(2019)Lemos, Lee, Efstathiou, \&
  Gratton}]{Lemos:2018smw}
Lemos, P., Lee, E., Efstathiou, G., \& Gratton, S. 2019, Mon. Not. Roy. Astron.
  Soc., 483, 4803, \dodoi{10.1093/mnras/sty3082}

\bibitem[{Lewis {et~al.}(2000)Lewis, Challinor, \& Lasenby}]{Lewis:1999bs}
Lewis, A., Challinor, A., \& Lasenby, A. 2000, Astrophys. J., 538, 473,
  \dodoi{10.1086/309179}

\bibitem[{Li \& Shafieloo(2019)}]{Li:2019yem}
Li, X., \& Shafieloo, A. 2019, Astrophys. J. Lett., 883, L3,
  \dodoi{10.3847/2041-8213/ab3e09}

\bibitem[{Li \& Shafieloo(2020)}]{Li:2020ybr}
---. 2020, Astrophys. J., 902, 58, \dodoi{10.3847/1538-4357/abb3d0}

\bibitem[{Li {et~al.}(2019)Li, Shafieloo, Sahni, \& Starobinsky}]{late-arman2}
Li, X., Shafieloo, A., Sahni, V., \& Starobinsky, A.~A. 2019, Astrophys. J.,
  887, 153, \dodoi{10.3847/1538-4357/ab535d}

\bibitem[{Lin {et~al.}(2019{\natexlab{a}})Lin, Benevento, Hu, \&
  Raveri}]{early-lin}
Lin, M.-X., Benevento, G., Hu, W., \& Raveri, M. 2019{\natexlab{a}}, Phys. Rev.
  D, 100, 063542, \dodoi{10.1103/PhysRevD.100.063542}

\bibitem[{Lin {et~al.}(2019{\natexlab{b}})Lin, Raveri, \& Hu}]{early-lin1}
Lin, M.-X., Raveri, M., \& Hu, W. 2019{\natexlab{b}}, Phys. Rev. D, 99, 043514,
  \dodoi{10.1103/PhysRevD.99.043514}

\bibitem[{Lue \& Starkman(2004)}]{Lue:2004za}
Lue, A., \& Starkman, G.~D. 2004, Phys. Rev. D, 70, 101501,
  \dodoi{10.1103/PhysRevD.70.101501}

\bibitem[{Maartens \& Koyama(2010)}]{Maartens:2010ar}
Maartens, R., \& Koyama, K. 2010, Living Rev. Rel., 13, 5,
  \dodoi{10.12942/lrr-2010-5}

\bibitem[{Mukohyama(2000)}]{Mukohyama:2000ui}
Mukohyama, S. 2000, Phys. Rev. D, 62, 084015,
  \dodoi{10.1103/PhysRevD.62.084015}

\bibitem[{Mukohyama(2001)}]{Mukohyama:2001yp}
---. 2001, Phys. Rev. D, 64, 064006, \dodoi{10.1103/PhysRevD.64.064006}

\bibitem[{{Okamatsu} {et~al.}(2021){Okamatsu}, {Sekiguchi}, \&
  {Takahashi}}]{Okamatsu:2021zao}
{Okamatsu}, F., {Sekiguchi}, T., \& {Takahashi}, T. 2021, \prd, 104, 023523,
  \dodoi{10.1103/PhysRevD.104.023523}

\bibitem[{Panpanich {et~al.}(2021)Panpanich, Burikham, Ponglertsakul, \&
  Tannukij}]{late-quintom}
Panpanich, S., Burikham, P., Ponglertsakul, S., \& Tannukij, L. 2021, Chin.
  Phys. C, 45, 015108, \dodoi{10.1088/1674-1137/abc537}

\bibitem[{Peirone {et~al.}(2019)Peirone, Benevento, Frusciante, \&
  Tsujikawa}]{late-shinji}
Peirone, S., Benevento, G., Frusciante, N., \& Tsujikawa, S. 2019, Phys. Rev.
  D, 100, 063540, \dodoi{10.1103/PhysRevD.100.063540}

\bibitem[{{Planck Collaboration} {et~al.}(2016){Planck Collaboration}, {Ade},
  {Aghanim}, {Arnaud}, {Ashdown}, {Aumont}, {Baccigalupi}, {Banday},
  {Barreiro}, {Bartlett}, {Bartolo}, {Battaner}, {Battye}, {Benabed},
  {Beno{\^\i}t}, {Benoit-L{\'e}vy}, {Bernard}, {Bersanelli}, {Bielewicz},
  {Bock}, {Bonaldi}, {Bonavera}, {Bond}, {Borrill}, {Bouchet}, {Boulanger},
  {Bucher}, {Burigana}, {Butler}, {Calabrese}, {Cardoso}, {Catalano},
  {Challinor}, {Chamballu}, {Chary}, {Chiang}, {Chluba}, {Christensen},
  {Church}, {Clements}, {Colombi}, {Colombo}, {Combet}, {Coulais}, {Crill},
  {Curto}, {Cuttaia}, {Danese}, {Davies}, {Davis}, {de Bernardis}, {de Rosa},
  {de Zotti}, {Delabrouille}, {D{\'e}sert}, {Di Valentino}, {Dickinson},
  {Diego}, {Dolag}, {Dole}, {Donzelli}, {Dor{\'e}}, {Douspis}, {Ducout},
  {Dunkley}, {Dupac}, {Efstathiou}, {Elsner}, {En{\ss}lin}, {Eriksen},
  {Farhang}, {Fergusson}, {Finelli}, {Forni}, {Frailis}, {Fraisse},
  {Franceschi}, {Frejsel}, {Galeotta}, {Galli}, {Ganga}, {Gauthier}, {Gerbino},
  {Ghosh}, {Giard}, {Giraud-H{\'e}raud}, {Giusarma}, {Gjerl{\o}w},
  {Gonz{\'a}lez-Nuevo}, {G{\'o}rski}, {Gratton}, {Gregorio}, {Gruppuso},
  {Gudmundsson}, {Hamann}, {Hansen}, {Hanson}, {Harrison}, {Helou},
  {Henrot-Versill{\'e}}, {Hern{\'a}ndez-Monteagudo}, {Herranz}, {Hildebrandt},
  {Hivon}, {Hobson}, {Holmes}, {Hornstrup}, {Hovest}, {Huang}, {Huffenberger},
  {Hurier}, {Jaffe}, {Jaffe}, {Jones}, {Juvela}, {Keih{\"a}nen}, {Keskitalo},
  {Kisner}, {Kneissl}, {Knoche}, {Knox}, {Kunz}, {Kurki-Suonio}, {Lagache},
  {L{\"a}hteenm{\"a}ki}, {Lamarre}, {Lasenby}, {Lattanzi}, {Lawrence}, {Leahy},
  {Leonardi}, {Lesgourgues}, {Levrier}, {Lewis}, {Liguori}, {Lilje},
  {Linden-V{\o}rnle}, {L{\'o}pez-Caniego}, {Lubin}, {Mac{\'\i}as-P{\'e}rez},
  {Maggio}, {Maino}, {Mandolesi}, {Mangilli}, {Marchini}, {Maris}, {Martin},
  {Martinelli}, {Mart{\'\i}nez-Gonz{\'a}lez}, {Masi}, {Matarrese}, {McGehee},
  {Meinhold}, {Melchiorri}, {Melin}, {Mendes}, {Mennella}, {Migliaccio},
  {Millea}, {Mitra}, {Miville-Desch{\^e}nes}, {Moneti}, {Montier}, {Morgante},
  {Mortlock}, {Moss}, {Munshi}, {Murphy}, {Naselsky}, {Nati}, {Natoli},
  {Netterfield}, {N{\o}rgaard-Nielsen}, {Noviello}, {Novikov}, {Novikov},
  {Oxborrow}, {Paci}, {Pagano}, {Pajot}, {Paladini}, {Paoletti}, {Partridge},
  {Pasian}, {Patanchon}, {Pearson}, {Perdereau}, {Perotto}, {Perrotta},
  {Pettorino}, {Piacentini}, {Piat}, {Pierpaoli}, {Pietrobon}, {Plaszczynski},
  {Pointecouteau}, {Polenta}, {Popa}, {Pratt}, {Pr{\'e}zeau}, {Prunet},
  {Puget}, {Rachen}, {Reach}, {Rebolo}, {Reinecke}, {Remazeilles}, {Renault},
  {Renzi}, {Ristorcelli}, {Rocha}, {Rosset}, {Rossetti}, {Roudier},
  {Rouill{\'e} d'Orfeuil}, {Rowan-Robinson}, {Rubi{\~n}o-Mart{\'\i}n},
  {Rusholme}, {Said}, {Salvatelli}, {Salvati}, {Sandri}, {Santos},
  {Savelainen}, {Savini}, {Scott}, {Seiffert}, {Serra}, {Shellard}, {Spencer},
  {Spinelli}, {Stolyarov}, {Stompor}, {Sudiwala}, {Sunyaev}, {Sutton},
  {Suur-Uski}, {Sygnet}, {Tauber}, {Terenzi}, {Toffolatti}, {Tomasi},
  {Tristram}, {Trombetti}, {Tucci}, {Tuovinen}, {T{\"u}rler}, {Umana},
  {Valenziano}, {Valiviita}, {Van Tent}, {Vielva}, {Villa}, {Wade}, {Wandelt},
  {Wehus}, {White}, {White}, {Wilkinson}, {Yvon}, {Zacchei}, \&
  {Zonca}}]{planck2016}
{Planck Collaboration}, {Ade}, P.~A.~R., {Aghanim}, N., {et~al.} 2016, \aap,
  594, A13, \dodoi{10.1051/0004-6361/201525830}

\bibitem[{{Planck Collaboration} {et~al.}(2020){Planck Collaboration},
  {Aghanim}, {Akrami}, {Ashdown}, {Aumont}, {Baccigalupi}, {Ballardini},
  {Banday}, {Barreiro}, {Bartolo}, {Basak}, {Battye}, {Benabed}, {Bernard},
  {Bersanelli}, {Bielewicz}, {Bock}, {Bond}, {Borrill}, {Bouchet}, {Boulanger},
  {Bucher}, {Burigana}, {Butler}, {Calabrese}, {Cardoso}, {Carron},
  {Challinor}, {Chiang}, {Chluba}, {Colombo}, {Combet}, {Contreras}, {Crill},
  {Cuttaia}, {de Bernardis}, {de Zotti}, {Delabrouille}, {Delouis}, {Di
  Valentino}, {Diego}, {Dor{\'e}}, {Douspis}, {Ducout}, {Dupac}, {Dusini},
  {Efstathiou}, {Elsner}, {En{\ss}lin}, {Eriksen}, {Fantaye}, {Farhang},
  {Fergusson}, {Fernandez-Cobos}, {Finelli}, {Forastieri}, {Frailis},
  {Fraisse}, {Franceschi}, {Frolov}, {Galeotta}, {Galli}, {Ganga},
  {G{\'e}nova-Santos}, {Gerbino}, {Ghosh}, {Gonz{\'a}lez-Nuevo}, {G{\'o}rski},
  {Gratton}, {Gruppuso}, {Gudmundsson}, {Hamann}, {Handley}, {Hansen},
  {Herranz}, {Hildebrandt}, {Hivon}, {Huang}, {Jaffe}, {Jones}, {Karakci},
  {Keih{\"a}nen}, {Keskitalo}, {Kiiveri}, {Kim}, {Kisner}, {Knox},
  {Krachmalnicoff}, {Kunz}, {Kurki-Suonio}, {Lagache}, {Lamarre}, {Lasenby},
  {Lattanzi}, {Lawrence}, {Le Jeune}, {Lemos}, {Lesgourgues}, {Levrier},
  {Lewis}, {Liguori}, {Lilje}, {Lilley}, {Lindholm}, {L{\'o}pez-Caniego},
  {Lubin}, {Ma}, {Mac{\'\i}as-P{\'e}rez}, {Maggio}, {Maino}, {Mandolesi},
  {Mangilli}, {Marcos-Caballero}, {Maris}, {Martin}, {Martinelli},
  {Mart{\'\i}nez-Gonz{\'a}lez}, {Matarrese}, {Mauri}, {McEwen}, {Meinhold},
  {Melchiorri}, {Mennella}, {Migliaccio}, {Millea}, {Mitra},
  {Miville-Desch{\^e}nes}, {Molinari}, {Montier}, {Morgante}, {Moss}, {Natoli},
  {N{\o}rgaard-Nielsen}, {Pagano}, {Paoletti}, {Partridge}, {Patanchon},
  {Peiris}, {Perrotta}, {Pettorino}, {Piacentini}, {Polastri}, {Polenta},
  {Puget}, {Rachen}, {Reinecke}, {Remazeilles}, {Renzi}, {Rocha}, {Rosset},
  {Roudier}, {Rubi{\~n}o-Mart{\'\i}n}, {Ruiz-Granados}, {Salvati}, {Sandri},
  {Savelainen}, {Scott}, {Shellard}, {Sirignano}, {Sirri}, {Spencer},
  {Sunyaev}, {Suur-Uski}, {Tauber}, {Tavagnacco}, {Tenti}, {Toffolatti},
  {Tomasi}, {Trombetti}, {Valenziano}, {Valiviita}, {Van Tent}, {Vibert},
  {Vielva}, {Villa}, {Vittorio}, {Wandelt}, {Wehus}, {White}, {White},
  {Zacchei}, \& {Zonca}}]{planck2018}
{Planck Collaboration}, {Aghanim}, N., {Akrami}, Y., {et~al.} 2020, \aap, 641,
  A6, \dodoi{10.1051/0004-6361/201833910}

\bibitem[{Poulin {et~al.}(2018)Poulin, Smith, Grin, Karwal, \&
  Kamionkowski}]{early-kamion1}
Poulin, V., Smith, T.~L., Grin, D., Karwal, T., \& Kamionkowski, M. 2018, Phys.
  Rev. D, 98, 083525, \dodoi{10.1103/PhysRevD.98.083525}

\bibitem[{Poulin {et~al.}(2019)Poulin, Smith, Karwal, \&
  Kamionkowski}]{early-kamion2}
Poulin, V., Smith, T.~L., Karwal, T., \& Kamionkowski, M. 2019, Phys. Rev.
  Lett., 122, 221301, \dodoi{10.1103/PhysRevLett.122.221301}

\bibitem[{Randall \& Sundrum(1999{\natexlab{a}})}]{Randall:1999ee}
Randall, L., \& Sundrum, R. 1999{\natexlab{a}}, Phys. Rev. Lett., 83, 3370,
  \dodoi{10.1103/PhysRevLett.83.3370}

\bibitem[{Randall \& Sundrum(1999{\natexlab{b}})}]{Randall:1999vf}
---. 1999{\natexlab{b}}, Phys. Rev. Lett., 83, 4690,
  \dodoi{10.1103/PhysRevLett.83.4690}

\bibitem[{Raveri(2020)}]{late-raveri19}
Raveri, M. 2020, Phys. Rev. D, 101, 083524, \dodoi{10.1103/PhysRevD.101.083524}

\bibitem[{Riess {et~al.}(2019)Riess, Casertano, Yuan, Macri, \&
  Scolnic}]{riess19}
Riess, A.~G., Casertano, S., Yuan, W., Macri, L.~M., \& Scolnic, D. 2019,
  Astrophys. J., 876, 85, \dodoi{10.3847/1538-4357/ab1422}

\bibitem[{{Riess} {et~al.}(2011){Riess}, {Macri}, {Casertano}, {Lampeitl},
  {Ferguson}, {Filippenko}, {Jha}, {Li}, \& {Chornock}}]{riess11A}
{Riess}, A.~G., {Macri}, L., {Casertano}, S., {et~al.} 2011, \apj, 730, 119,
  \dodoi{10.1088/0004-637X/730/2/119}

\bibitem[{{Riess} {et~al.}(2016){Riess}, {Macri}, {Hoffmann}, {Scolnic},
  {Casertano}, {Filippenko}, {Tucker}, {Reid}, {Jones}, {Silverman},
  {Chornock}, {Challis}, {Yuan}, {Brown}, \& {Foley}}]{riess16}
{Riess}, A.~G., {Macri}, L.~M., {Hoffmann}, S.~L., {et~al.} 2016, \apj, 826,
  56, \dodoi{10.3847/0004-637X/826/1/56}

\bibitem[{{Riess} {et~al.}(2018){Riess}, {Casertano}, {Yuan}, {Macri},
  {Bucciarelli}, {Lattanzi}, {MacKenty}, {Bowers}, {Zheng}, {Filippenko},
  {Huang}, \& {Anderson}}]{riess18}
{Riess}, A.~G., {Casertano}, S., {Yuan}, W., {et~al.} 2018, \apj, 861, 126,
  \dodoi{10.3847/1538-4357/aac82e}

\bibitem[{Rossi {et~al.}(2019)Rossi, Ballardini, Braglia, Finelli, Paoletti,
  Starobinsky, \& Umilt\`a}]{early-starobinsky}
Rossi, M., Ballardini, M., Braglia, M., {et~al.} 2019, Phys. Rev. D, 100,
  103524, \dodoi{10.1103/PhysRevD.100.103524}

\bibitem[{Sahni \& Shtanov(2003)}]{ss02}
Sahni, V., \& Shtanov, Y. 2003, JCAP, 11, 014,
  \dodoi{10.1088/1475-7516/2003/11/014}

\bibitem[{Sahni \& Shtanov(2005)}]{loiter}
---. 2005, Phys. Rev. D, 71, 084018, \dodoi{10.1103/PhysRevD.71.084018}

\bibitem[{Sahni {et~al.}(2005)Sahni, Shtanov, \& Viznyuk}]{mimicry}
Sahni, V., Shtanov, Y., \& Viznyuk, A. 2005, JCAP, 12, 005,
  \dodoi{10.1088/1475-7516/2005/12/005}

\bibitem[{Sahni \& Starobinsky(2006)}]{Sahni:2006pa}
Sahni, V., \& Starobinsky, A. 2006, Int. J. Mod. Phys. D, 15, 2105,
  \dodoi{10.1142/S0218271806009704}

\bibitem[{Sawicki {et~al.}(2007)Sawicki, Song, \& Hu}]{Sawicki:2006jj}
Sawicki, I., Song, Y.-S., \& Hu, W. 2007, Phys. Rev. D, 75, 064002,
  \dodoi{10.1103/PhysRevD.75.064002}

\bibitem[{Schmidt(2009)}]{Schmidt:2009sv}
Schmidt, F. 2009, Phys. Rev. D, 80, 123003, \dodoi{10.1103/PhysRevD.80.123003}

\bibitem[{Seahra \& Hu(2010)}]{Seahra:2010fj}
Seahra, S.~S., \& Hu, W. 2010, Phys. Rev. D, 82, 124015,
  \dodoi{10.1103/PhysRevD.82.124015}

\bibitem[{Seto \& Toda(2021)}]{Seto:2021xua}
Seto, O., \& Toda, Y. 2021, Phys. Rev. D, 103, 123501,
  \dodoi{10.1103/PhysRevD.103.123501}

\bibitem[{Shafieloo {et~al.}(2018)Shafieloo, Hazra, Sahni, \&
  Starobinsky}]{late-arman1}
Shafieloo, A., Hazra, D.~K., Sahni, V., \& Starobinsky, A.~A. 2018, Mon. Not.
  Roy. Astron. Soc., 473, 2760, \dodoi{10.1093/mnras/stx2481}

\bibitem[{Shtanov(2000)}]{Shtanov:2000vr}
Shtanov, Y.~V. 2000.
\newblock \doarXiv{hep-th/0005193}

\bibitem[{Smith {et~al.}(2020)Smith, Poulin, \& Amin}]{early-smith}
Smith, T.~L., Poulin, V., \& Amin, M.~A. 2020, Phys. Rev. D, 101, 063523,
  \dodoi{10.1103/PhysRevD.101.063523}

\bibitem[{Vagnozzi(2020)}]{late-vagnozzi}
Vagnozzi, S. 2020, Phys. Rev. D, 102, 023518,
  \dodoi{10.1103/PhysRevD.102.023518}

\bibitem[{{Vagnozzi}(2021)}]{Vagnozzi:2021gjh}
{Vagnozzi}, S. 2021, \prd, 104, 063524, \dodoi{10.1103/PhysRevD.104.063524}

\bibitem[{Vagnozzi {et~al.}(2021)Vagnozzi, Pacucci, \& Loeb}]{Vagnozzi:2021tjv}
Vagnozzi, S., Pacucci, F., \& Loeb, A. 2021.
\newblock \doarXiv{2105.10421}

\bibitem[{Viznyuk {et~al.}(2018)Viznyuk, Bag, Shtanov, \&
  Sahni}]{Viznyuk:2018eiz}
Viznyuk, A., Bag, S., Shtanov, Y., \& Sahni, V. 2018, Phys. Rev. D, 98, 064024,
  \dodoi{10.1103/PhysRevD.98.064024}

\bibitem[{Viznyuk {et~al.}(2014)Viznyuk, Shtanov, \& Sahni}]{Viznyuk:2013ywa}
Viznyuk, A., Shtanov, Y., \& Sahni, V. 2014, Phys. Rev. D, 89, 083523,
  \dodoi{10.1103/PhysRevD.89.083523}

\bibitem[{Ye \& Piao(2020)}]{Ye:2020btb}
Ye, G., \& Piao, Y.-S. 2020, Phys. Rev. D, 101, 083507,
  \dodoi{10.1103/PhysRevD.101.083507}

\end{thebibliography}
\end{document}